\documentclass[preprint2]{aastex}

\usepackage{amsmath}
\usepackage{graphicx}
\usepackage{epsfig,psfrag,epic,eepic}
\usepackage{textcomp}
\usepackage{times}
\usepackage{longtable}

\slugcomment{Submitted to the Astrophysical Journal}

\shorttitle{Multiplicity in Sco-Cen}
\shortauthors{Janson et al.}

\begin{document}

\title{A Multiplicity Census of Intermediate-Mass Stars in Scorpius-Centaurus\altaffilmark{*}}

\author{Markus Janson\altaffilmark{1,8}, 
David Lafreni{\`e}re\altaffilmark{2}, 
Ray Jayawardhana\altaffilmark{3}, 
Mariangela Bonavita\altaffilmark{3,4}, 
Julien H. Girard\altaffilmark{5}, 
Alexis Brandeker\altaffilmark{6}, 
John E. Gizis\altaffilmark{7}
}

\altaffiltext{*}{Based on Gemini observations from programs GS-2011A-Q-44, GS-2012A-Q-18, GS-2012A-DD-6, GS-2013A-Q-21, and on ESO observations from program 089.C-0422(A).}
\altaffiltext{1}{Department of Astrophysical Sciences, Princeton University, Princeton, NJ, USA; \texttt{janson@astro.princeton.edu}}
\altaffiltext{2}{Department of Physics, University of Montreal, Montreal, QC, Canada}
\altaffiltext{3}{Department of Astronomy and Astrophysics, University of Toronto, Toronto, ON, Canada}
\altaffiltext{4}{Osservatorio Astronomico di Padova - INAF, Padova, Italy}
\altaffiltext{5}{European Southern Observatory, Santiago, Chile}
\altaffiltext{6}{Department of Astronomy, Stockholm University, Stockholm, Sweden}
\altaffiltext{7}{Department of Physics and Astronomy, University of Delaware, Newark, DE, USA}
\altaffiltext{8}{Hubble fellow}

\begin{abstract}\noindent
Stellar multiplicity properties have been studied for much of the range from the lowest to the highest stellar masses, but intermediate-mass stars from F-type to late A-type have received relatively little attention. Here we report on a Gemini/NICI snapshot imaging survey of 138 such stars in the young Scorpius-Centaurus (Sco-Cen) region, for the purpose of studying multiplicity with sensitivity down to planetary masses at wide separations. In addition to two brown dwarfs and a companion straddling the hydrogen burning limit we reported previously, here we present 26 new stellar companions and determine a multiplicity fraction within 0.1\arcsec--5.0\arcsec\ of 21$\pm$4\%. Depending on the adopted semi-major axis distribution, our results imply a total multiplicity in the range of $\sim$60--80\%, which further supports the known trend of a smoothly continuous increase in the multiplicity fraction as a function of primary stellar mass. A surprising feature in the sample is a distinct lack of nearly equal-mass binaries, for which we discuss possible reasons. The survey yielded no additional companions below or near the deuterium-burning limit, implying that their frequency at $>$200~AU separations is not quite as high as might be inferred from previous detections of such objects within the Sco-Cen region.
\end{abstract}


\section{Introduction}

The Scorpius-Centaurus (Sco-Cen) region is a young ($\sim$5--10~Myr) and relatively nearby ($\sim$120--150~pc) stellar association \citep{dezeeuw1999}, consisting of the sub-regions Upper Scorpius (USco), Upper Centaurus Lupus (UCL), and Lower Centaurus Crux (LCC). Given its young age, in particular for USco which is the youngest sub-region, it is a promising target for direct imaging searches for wide planetary companions. Several such surveys have been conducted in USco, and have led to a surprisingly large number of detections \citep{lafreniere2008a,ireland2011,lafreniere2011}, which has in turn led to the tentative estimation that as many as 4$\%$ of stars may have very wide ($>$200~AU) high-mass planets or very low-mass brown dwarfs \citep{ireland2011}. Sco-Cen has also been a favourable target region for multiplicity studies \citep[e.g.][]{shatsky2002,kouwenhoven2007}. Some of the reasons for this are the uniformity of the region in distance and age, and the high completeness that can be reached down to low companion masses. 

It is well known that multiplicity properties depend on the mass of the primary star \citep[e.g.][]{duchene2013}, and extensive multiplicity surveys have been performed over several different mass ranges in recent years \citep[e.g.][]{kouwenhoven2007,raghavan2010,janson2012a}. However, intermediate-mass stars in the range of $\sim$1--3~$M_{\rm sun}$ has received relatively little attention in this regard. Hence, it is an interesting range for testing the consistency and continuity of dependencies in multiplicity properties on stellar mass. It is also an interesting mass range from the point of view of exoplanet imaging, since several planets and low-mass substellar companions have been imaged around such primaries \citep{marois2008,lagrange2009,carson2013}, which has motivated targeted surveys of intermediate-mass stars in the recent past \citep{janson2011,vigan2012}.

Motivated by these issues, we have performed a snapshot imaging survey of 138 Sco-Cen stars with spectral types primarily in the F-type and late A-type range that have not been previously observed in a high-contrast context, to assess stellar multiplicity with a high completeness down through the brown dwarf range, and sensitivity to planetary masses at wide separations. The survey is performed in the context of other studies of multiplicity in young stellar associations that we are performing or have recently performed, including one survey in Chameleon I \citep{lafreniere2008b}, one in the Taurus star-forming region (Daemgen et al., in prep.), and one specifically in USco (Lafreni{\`e}re et al., in prep.). Together, these surveys will span an age range of $\sim$1--10~Myr.

The paper is structured in the following way: In Sect. \ref{s:odr}, we describe the observational aspects of the study, with a description of the sample in Sect. \ref{s:sample}, the observational setup in Sect. \ref{s:observations}, and the data reduction procedure in Sect. \ref{s:reduction}. The results and their analysis are then presented in Sect. \ref{s:results}, including the astrometric analysis in Sect. \ref{s:astrometry}, the determination of companion properties in Sect. \ref{s:companions}, the completeness estimation in Sect. \ref{s:completeness}, statistical properties in Sect. \ref{s:statistics} and individual notes for particular targets in Sect. \ref{s:notes}. Finally, we discuss our results in a broader context in Sect. \ref{s:discussion}.

\section{Observations and Data Reduction}
\label{s:odr}

\subsection{Sample Selection}
\label{s:sample}

Our sample consists of early-type stars in the Sco-Cen region. It was selected among all targets identified as Sco-Cen members in \citep{dezeeuw1999} that fulfilled two criteria: 1) They had to have a measured parallax and proper motion by Hipparcos, and 2) they had to have been previously unobserved with adaptive optics (AO) instruments on large telescopes; i.e., in the surveys of \citet{shatsky2002} or \citet{kouwenhoven2007}, or in our NIRI survey of the USco region (Lafreni{\`e}re et al., in prep.). The Hipparcos-based requirement allowed for automatic selection of targets with well-determined kinematics and distances, which is important both for a high fidelity in selection of bona fide members, as well as being helpful for the physical interpretation of any discovered companions. Indeed, the vast majority of early-type Sco-Cen stars that were followed up in \citet{chen2011} were confirmed as real members, whereas later-type stars had a lower confirmation rate. Likewise, all our targets that have been studied in \citet{rizzuto2011} were assigned high membership probabilities. The visual brightness limit of Hipparcos of $V \sim 9$~mag also matches well with the range of optimal performance for AO wave front sensors, and thus ensures that a good contrast can be achieved in each case (provided acceptable ambient conditions). The avoidance of \citet{shatsky2002} and \citet{kouwenhoven2007} targets helped both to ensure a maximal utility of our survey in terms of mapping the full multiplicity properties of Sco-Cen, and also honed in on an interesting range in planet properties -- since the previous surveys focused largely on more massive stars (B- and early A-type), the remaining targets are primarily F- and late A-type (with a few cases of early G-type), which is a range that has not been covered as extensively in multiplicity studies as most other spectral type ranges thus far. 

As a result of these selections, our total sample consisted of 145 stars of which 138 were eventually observed, which have a median mass of 1.5~$M_{\rm sun}$ and range from 1.0~$M_{\rm sun}$ to 4.2~$M_{\rm sun}$ (mass determinations are described in Sect. \ref{s:companions}). While 4 stars have masses of $>$5~$M_{\rm sun}$, they were imaged with a different instrumental setting that disfavours a homogenous analysis, and by chance several of them were also taken under rather poor conditions, hence they have simply been excluded from any statistical analysis. The ages of the targets adopted for the analysis performed here are 5~Myr for the (relatively few) USco members and 10~Myr for the UCL and LCC members. These ages have been under discussion in the recent literature, with \citet{pecaut2012} suggesting an older age, but we base our estimates on the \citet{song2012} analysis, which empirically demonstrates UCL and LCC to be younger than $\beta$~Pic, adopting an age of 10 Myr in both cases. The USco region is not discussed in \citet{song2012}, but since it is known to be younger than UCL and LCC, we adopt the original age of 5~Myr \citep{dezeeuw1999}. There are two scientific issues studied here that are in principle impacted by the age: The detection limit estimation, and the mass ratio determinations. However, as we will discuss in the individual sections, the impact of a factor $\sim$2 change in age would be modest for the purposes of this study. The targets are summarized in table \ref{t:sample}.

\subsection{Observational Procedure}
\label{s:observations}

The first epoch imaging was performed in the Spring of 2011, using the NICI AO-assisted camera \citep{artigau2008} at the Gemini South telescope in Chile. Out of 145 proposed targets, 115 were observed during this period. Follow-up of 53 targets with companion candidates was executed with the same instrument one year later, during observing cycle 2012A. Furthermore, due to an unforseen excess in available observing time with NICI, our 2011A program was re-introduced into the queue in 2012A, and an additional 23 targets were imaged, for a total of 138 targets observed in at least one epoch. Some candidates that had indications of sharing a common proper motion with the primary were followed up spectroscopically, either using VLT/NACO \citep{lenzen2003,rousset2003} during ESO period 89 for simultaneous H+K coverage with AO-assisted slit spectroscopy, or using Gemini/NIFS \citep{mcgregor2002} during 2012A for AO-assisted H-band integral field spectroscopy. Some targets were also followed up in a third astrometric epoch using excess time for the 2012A program that arose from efficient scheduling and execution of the observations. Finally, in period 2013A we performed follow-up of those targets with candidates that had only been observed in one epoch during 2012A, in addition to acquiring another epoch of imaging for the particularly puzzling targets HIP~80130 and HIP~82569.

Our imaging observations were optimized for a high observing efficiency, and consisted of two sequential exposures per target, where the first exposure included 10 coadds of 0.38 seconds each, which is the minimal individual integration time available for NICI. The second exposure consisted of a single 80 second integration. The dual band imaging mode was employed with the 50/50 beamsplitter, using the $K_{\rm s}$ filter in the red channel and the $H_2$(1-0) filter in the blue channel. The $H_2$(1-0) filter is a narrow-band filter within the $K$-band range, with almost identical pivotal wavelength (2.12~$\mu$m) to the $K_{\rm s}$ filter. In this way, we achieved a nearly simultaneous and very wide dynamic range, with the primary star being generally non-saturated in the short $H_2$(1-0) exposures, and with good sensitivity with respect to the read noise limit for faint candidates in the long $K_{\rm s}$ exposures. Each target was subjected to a random offset of $<$5\arcsec\ after acquisition. This allowed adjacently observed targets to be used as sky frames for each other, in the same way as during dithering/jittering, but applied on a sequence of targets instead of a sequence of images of a single target. The observations were generally acquired under average to good conditions (fulfilling the Gemini 70 percentile image quality condition), with a few exceptions that are marked in Table \ref{t:sample} and excluded from any statistical analysis.

The spectroscopic observations were mainly utilized for the analysis presented in \citep{janson2012b}, but were also useful for some of the targets presented specifically here. Summarizing the description in the previous work, the NACO spectra were taken in the $H+K$ setting with a spectral resolution of $R \sim 550$ and a simultaneous coverage from 1.33 to 2.53~$\mu$m. An ABBA nodding along the 172~mas wide slit was employed for background subtraction, with a nod throw of 10\arcsec. One initial AB cycle was taken with short integration times of 1~s and 12 coadds, followed by a series of 100~s exposures, structured as 3 AB cycles with 5~s by 20 coadds for the brightest candidates and 10 AB cycles with 25~s by 4 coadds for the faintest. The short exposures allow for the primary to remain non-saturated such that it can be used as a reference star, and the long exposures minimize the read noise allowing for a high $S/N$ of the companion candidate. The NIFS spectra were taken in $H$-band with a spectral resolution of $R \sim 5000$ covering a spectral range of 1.49 to 1.80~$\mu$m. Each sequence observation started with a dither sequence with 21~s integration times with the central star in the field of view, to get non-saturated spectra of the primary for the purpose of using it as a standard star. This was then followed by a deeper dithering sequence with integration times of 240~s, with the companion candidate included in the field of view. If the separation of the candidate was small enough that the NIFS field of view (3\arcsec\ on each side) could fit both the star and companion then this was accommodated, otherwise the field was centered on the candidate.

\subsection{Data Reduction}
\label{s:reduction}

Data reduction for the NICI imaging was performed using custom IDL routines, since the observational strategy was somewhat novel and thus required flexibility in the reduction procedure. For each image file, the red and blue channel images were extracted and analyzed separately. As a first step, flat fielding and bad pixel removal were applied, followed by a background subtraction, in which a median of several adjacent images were taken and subtracted from each individual image. Distortion correction was applied using the separate distortion solutions for the red and blue channel provided on the NICI homepage\footnote{\url{http://www.gemini.edu/?q=node/10493}}, with a quadratic interpolation scheme. By default, North points downward in NICI images, and the red and blue channels are mirror images of each other, with one having a right-handed and the other a left-handed orientation. Which channel is oriented which way depends on whether NICI is mounted in an up-looking or side-looking configuration. Thus, since the instrument was mounted in different configurations at different epochs, the orientation switched between the red and blue channels between images. All images were re-oriented into a common framework with North pointing up and East to the right, taking these various circumstances into account. Finally, the images were shifted using spline interpolation, such that the primary star became centered on the central pixel. For this purpose, Gaussian centroiding was used for the $H_2$(1-0) images, where the star is unsaturated in the short exposures. For the $K_{\rm s}$ images, where the primary is typically saturated even in the short exposures, manual centering by eye was applied. As described in \citet{janson2012b}, this gives a smaller scatter in the background star astrometry than a range of other methods tested for the purpose, and the good consistency between the independent $H_2$(1-0) and $K_{\rm s}$ astrometry noted in Sect. \ref{s:astrometry} further demonstrates the validity of the method.

In the same way as for the NICI imaging, the NACO spectroscopy data reduction was also performed with a custom IDL pipeline, since we had a set of routines available from a previous observing program \citep{janson2010} which could be easily adapted to form part of the pipeline. The procedure started with flat fielding and bad pixel removal. Each AB set was then pairwise subtracted to remove the background. The spectral traces are (nearly) vertical on the NACO detector, hence for each pixel row, the photocenter of the star was determined through Gaussian centroiding. This works well not only for the non-saturated sequences, but also for the longer-exposure sequences, since the stars are only mildly saturated. A spectral trace was then fitted to the centers of the respective rows, and all data were shifted so as to form a perfectly vertical spectral trace, with the photocenter of the star at the central pixel column, for all frames in each observational sequence. One collapsed frame of the collected non-saturated data and one of the saturated data were produced using a regular mean combination. At this point, the secondary spectra were clearly visible at their known positions in the deep exposures, and could be extracted using an interpolation between the fluxes measured directly inside and outside of the location of the secondary as an estimation of the stellar PSF at that location. A 162~mas aperture was used in the extraction of both the stellar and companion spectra. Wavelength calibration was performed using a combination of telluric features and intrinsic features in the stellar spectra. For flux calibration, the primary star was modeled as a single-temperature blackbody. The extracted companion spectrum was divided by the fraction of the stellar spectrum to the model blackbody, which eliminates all telluric features from the companion's spectrum. Intrinsic stellar features remain as contaminants, but as noted in \citet{janson2012b}, such features are rare and weak in these early-type stars, and do not affect the largely continuum-based analysis that they are used for.

Basic data reduction of the NIFS data was done using the facility-provided IRAF pipeline. This executed all fundamental steps such as flat field correction, distortion correction, wavelength calibration, and data cube construction. The final steps of registering and shifting all frames to a common center, as well as extracting the spectra, were done in IDL. Each wavelength slice of each data cube was treated as an individual image, being centroided and interpolated in the same way as described for the imaging above. Extraction was performed using a 172~mas circular aperture.

\section{Analysis and Results}
\label{s:results}

\begin{figure}[p]
\centering
\includegraphics[width=8cm]{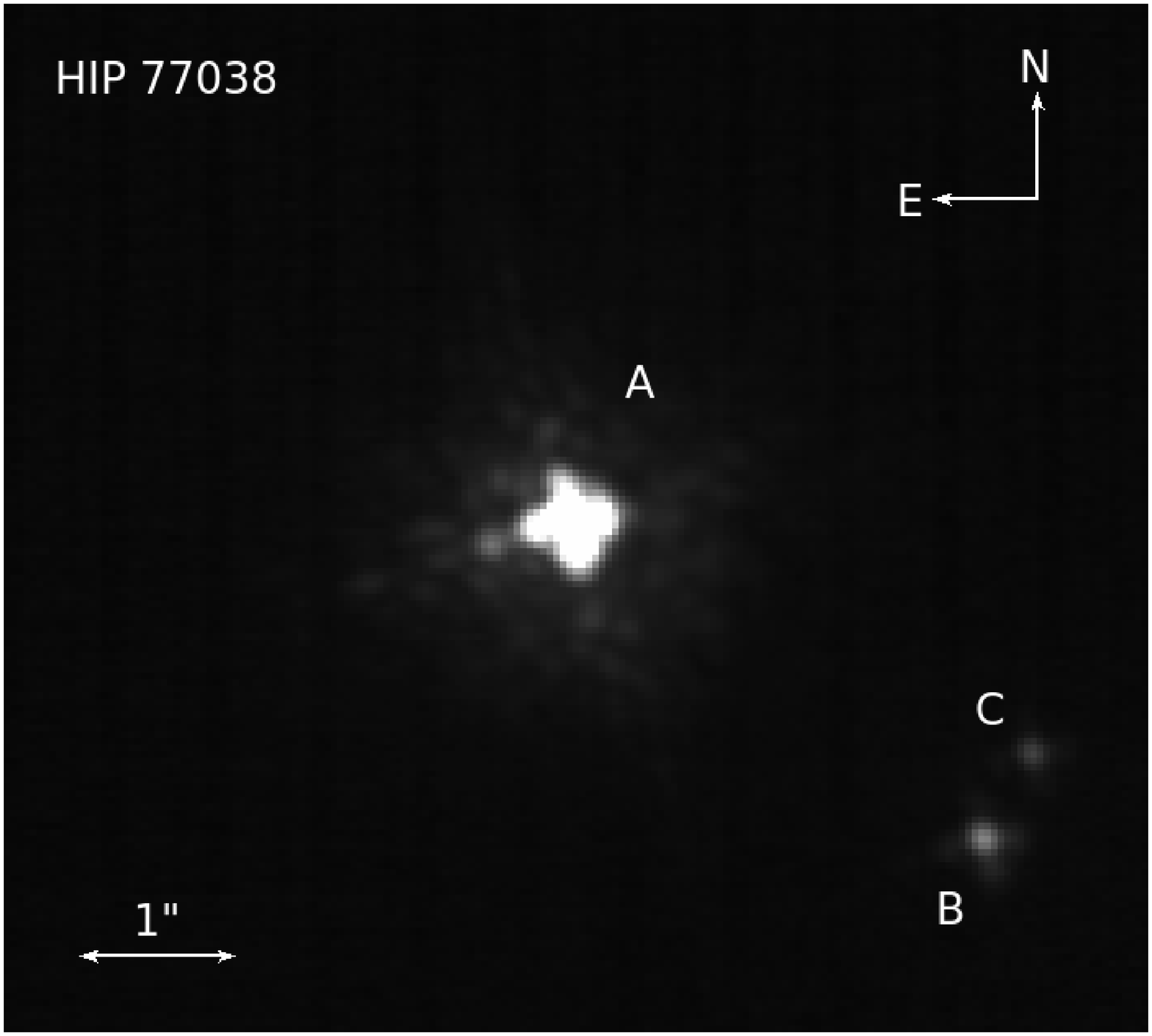}
\caption{Example image from the survey, showing the triple system HIP~77038. The tertiary component is a very low-mass star, with an estimated mass of $\sim$90~$M_{\rm jup}$. The apparent point source to the East of the primary at a small separation is a known ghost feature.}
\label{f:im77038}
\end{figure}

In this section, we describe the various analyses that were applied to the companion candidates detected in the images, and their results. The candidates considered for analysis here are exclusively those that have a projected separation between 0.1\arcsec and 5.0\arcsec from their parent stars. There are two reasons for the outer limit: Firstly, while the NICI field of view is 18\arcsec on each side, the dithering scheme with different stars being placed at different parts of the detector means that the completeness drops rapidly outside of 5\arcsec. Secondly, the false positive rate scales with the square of the angular separation, such that essentially all targets have false positives at $>$5\arcsec, and so follow-up of such very wide candidates becomes observationally inefficient. In total, we have discovered 145 candidates around 79 stars. Of these, 116 are considered either indicated or confirmed background sources, and 29 are considered indicated or confirmed companions (residing in 27 systems, since two systems are triple, see Fig. \ref{f:im77038} and Sect. \ref{s:notes}). To the best of our knowledge, none of the companions have been previously reported in the literature. In addition, there are three cases (HIP~63692, HIP~66001, and HIP~79097) where the PSF of the primary star is extended, which might point to the existence of a partially resolved close companion well inside of 0.1\arcsec, and there are two cases (HIP~50847 and HIP~64322) in which a probable binary companion is seen but the images are among those that were taken in too poor conditions for any solid conclusion to be drawn. These individual cases are discussed in Sect. \ref{s:notes}.

One of the grounds for assessment of companionship was based on calculated false alarm probabilities of individual candidates. These were estimated in the same way as in \citet{lafreniere2008b}, on the basis of the brightness of the candidate, its separation from the primary star, and the background stellar surface density at its location in the sky. The latter was acquired from 2MASS \citep{skrutskie2006} point source counts within a 15\arcmin\ radius from the primary star. For a given candidate, we then calculated the number of stars at least as bright as the candidate, which would fall within a circular area out to the separation of the candidate. In this way, a candidate whose properties are reproduced by, e.g., 0.01 background contaminants can be said to have a 1\% false alarm probability. It is important to interpret such a number within the context of the full survey -- i.e., a 1\% probability may seem small, but in a survey such as this one with $>$100 targets, an occurrence of one such contaminant is entirely plausible.

\subsection{Astrometry of Companions and Candidates}
\label{s:astrometry}

As a general broad classification, the candidate companions can be divided into a bright group and a faint group. The brighter candidates generally have very small background contamination probabilities ($\ll$1\%), favourably small separations, estimated masses in the stellar regime, and are visible in both the $K_{\rm s}$ and the $H_2$(1-0) images. The fainter candidates, by contrast, generally have high contamination probabilities ($\gg$1\%), favourably large separations, estimated masses in the planetary regime if they would be interpreted as companions ($\sim$5--15~$M_{\rm jup}$), and are generally not visible in the $H_2$(1-0) images. There are a few intermediate cases, some of which were discussed in \citet{janson2012b} and the rest of which will be discussed individually here, but first we will discuss the separate analyses that were applied to the two distinct populations.

The brighter candidates, as mentioned, were visible in both $K_{\rm s}$ and $H_2$(1-0) images, and so astrometric analysis was performed independently in the two channels for these candidates. We used Gaussian centroiding for determining the positions of the companions in both cases. The astrometry was found to be well consistent between the two bands, except for a systematic rotational shift of $1.17^{\rm o}$ in the $H_2$(1-0) data with respect to the $K_{\rm s}$ data, which switches sign between the up-looking and side-looking instrumental configurations. Hence, we interpret the blue channel as having a $1.17^{\rm o}$ rotational offset that is corrected for by de-rotating the images by the corresponding amount. We then evaluate errors in the astrometry on the basis of the scatter between the $H_2$(1-0) and the $K_{\rm s}$ astrometry. We do this separately for the very close ($\leq$0.5\arcsec) and the wider (0.5\arcsec--5\arcsec) population of candidates, and find errors of 4~mas in separation and $0.9^{\rm o}$ in position angle in the former case, and 8~mas and $0.4^{\rm o}$ respectively in the latter.

Given that these objects have very low contamination probabilities and increase in frequency toward smaller angular separations in an opposite way from what would be expected for background stars, they are considered to be probable companions and are considered as such for the remainder of the discussion. They have not been confirmed with common proper motion, except in cases where a fainter candidate was followed up in the same system. Among the targets that were followed up for this reason, HIP~58220~B can be confirmed as a common proper motion companion. HIP~57595~B could not be accurately registered in the second epoch due to overlap with a bright sidelobe from the primary star, although by eye it does appear to share a common proper motion. HIP~75891~B and HIP~77520~B are consistent with common proper motion, but the background hypothesis cannot be rejected due to a small motion of the primary stars relative to the astrometric precision. One exception, however, is HIP~72630, where the candidate has a false alarm probability of only 0.3\%, but the proper motion analysis nonetheless shows that it is a background contaminant. The candidate has the highest false alarm probability of all the targets with $\ll$1\%, hence while its presence is relatively unlikely, it is not particularly surprising. It does however demonstrate that common proper motion testing would be an important aspect of providing final proof of companionship for the individual binaries, and to weed out any potential contaminant that could conceivably remain in the sample, e.g. HIP~58528~B (false alarm probability of 0.2\%) or HIP~67428 (0.3\%).

The fainter candidates are generally not visible in the $H_2$(1-0) images, but as noted above, the astrometry in $K_{\rm s}$ and $H_2$(1-0) are well consistent, hence we proceed with astrometry on $K_{\rm s}$ alone for this population. It is known a priori that the vast majority of these candidates must be background contaminants from the calculated probabilities alone, so CPM (Common Proper Motion) analysis is certainly necessary in order to detect any real companions among them. Thus, all of the faint candidates have been followed up in at least one additional epoch. Since Sco-Cen targets move rather slowly on the sky, the confidence level of the common proper motion testing rarely reaches 3$\sigma$ or higher. However, the candidates move by a median amount of 27~mas, larger than the residual scatter of 13~mas, and the median difference between the candidate motion and the expected background trajectories is only 4~mas. Hence, the typical faint candidate shows a motion that is well consistent with background motion, and is distinct from common proper motion by just above the 2$\sigma$ level.

Nonetheless, at these levels of confidence, both real companions and background contaminants can plausibly exist among the candidates that experience relatively little motion. Hence, for the target that seemed to move the least or sometimes not at all, we performed further follow-up in a third epoch of astrometry, as well as with spectroscopy in many cases. In this way, the brown dwarfs discussed in \citet{janson2012b} were confirmed as companions (HIP~65423~B, HIP~65517~B, and HIP~72099~B), both through the CPM and spectral analyses consistently. Similarly, most of the rest of the targets could be confirmed as background contaminants in the CPM analysis, and in the cases where spectra were acquired (e.g. for the candidates around HIP~62677 and HIP~72584) a consistent result was acquired. However, there are two cases that remain puzzling in this regard. These targets -- HIP~80130 and HIP~82569 -- will be discussed in more detail in Sect. \ref{s:notes}. The multi-epoch astrometry is summarized in Table~\ref{t:astro}.

\subsection{Companion Properties}
\label{s:companions}

In order to determine the physical properties of the binaries discovered in our sample, we first calculate their photometric properties. The secondaries in the systems are typically clearly visible and non-saturated in both the short $H_2$(1-0) and the short $K_{\rm s}$ exposures. The primaries, on the other hand, are typically only non-saturated in the $H_2$(1-0) filter. However, the pivotal wavelengths of the respective filters are almost identical (2.12~$\mu$m in both cases), which means that the [$H_2 - K_{\rm s}$] color is essentially independent of stellar temperature. Indeed, integration of the flux in the bandpasses of the respective filters shows that the difference in this color between a 5000~K and a 10000~K blackbody is less than 0.01~mag. This temperature range encompasses all of our primary stars as well as Vega, implying that in Vega magnitudes, the $H_2$(1-0) and $K_{\rm s}$ magnitudes are essentially identical. As a result, we can estimate the primary flux count in the $K_{\rm s}$ image by multiplying the measured count in the $H_2$(1-0) image with a uniform factor. We determine this factor by measuring the fluxes of all secondaries that are non-saturated in both images, and find that the number is 14.02. Hence, we can estimate the $K_{\rm s}$-band binary flux ratios by estimating the primary counts in this way and by measuring the secondary counts directly. In addition, we can measure the $H_2$(1-0)-band binary flux ratios directly in the image, and by comparing the ratios we can cross-validate the method, and get an estimate of the errors associated with it. We find that the two measures are consistent to within 0.2~mag, which we use as the photometric error. In all cases, we use a circular aperture of 72~mas diameter to determine the counts.

For calculating masses of the various components, apparent $K$-band magnitudes are adopted from 2MASS \citep{skrutskie2006}, and the primary and secondary magnitudes are calculated using the flux ratios described above. The apparent magnitudes are translated into absolute magnitudes using distance moduli based on the Hipparcos parallaxes \citep{perryman1997}. This is then used in conjunction with isochrones at the ages of the Sco-Cen sub-groups to derive masses. In our case, we use the \citet{siess2000} isochrones, since they cover the full range of stellar masses in our sample from 0.1~$M_{\rm sun}$ to 5~$M_{\rm sun}$. However, we also compare the results to the \citet{baraffe1998} isochrones in the overlapping range of $\leq$1.4~$M_{\rm sun}$ and find that at these ages and masses, the predicted $K$-band magnitudes are consistent to within 0.1~mag between the models, which is smaller then the photometric error. We also use the \citet{baraffe1998} isochrones for one secondary with a mass of $<$100$M_{\rm jup}$, i.e. outside of the range of the \citet{siess2000} models. The magnitudes and masses of the binary components are listed in Table \ref{t:binary}.

\subsection{Detection Limits}
\label{s:completeness}

\begin{figure}[p]
\centering
\includegraphics[width=8cm]{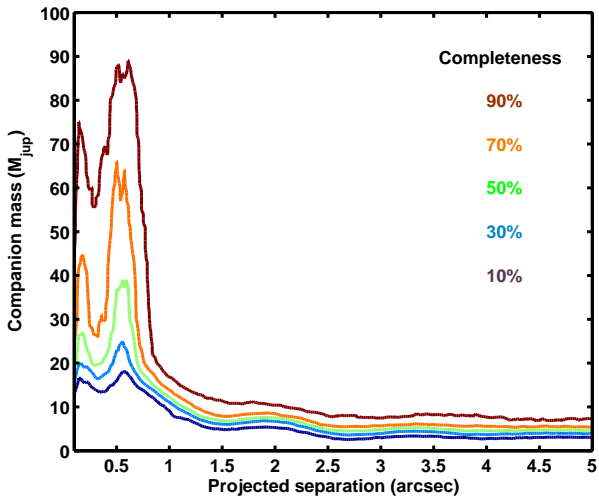}
\caption{Countours of the survey completeness as a function of separation and companion mass. The observations are essentially fully complete to stellar companions within 0.1\arcsec--5.0\arcsec, and provide good completeness down to planetary masses at wide separations.}
\label{f:completeness}
\end{figure}

For evaluating the detectability of companions around any given star in the survey, and for evaluating the completeness to such companions in the survey as a whole, we must first calculate contrast curves. This is done in a very similar manner to the contrast calculation of actual binaries described in Sect. \ref{s:companions}, by estimating the primary flux count in the deep $K_{\rm s}$-band images from the non-saturated $H_2$(1-0) count and factoring in the filter translation, as well as the difference in integration times. The standard deviation is calculated in a series of annuli at different separations from the central star to acquire $\sigma$ as a function of angular separation, and a 5$\sigma$ criterion is used as the detection threshold. Normalizing the 5$\sigma$ curves by the primary flux gives the contrast curve, and factoring in the primary magnitude and the distance modulus as in Sect. \ref{s:companions} gives the detectable absolute magnitude as function of separation around each star. Since these limits correspond to planet and brown dwarf masses, we use DUSTY \citep{chabrier2000} models for temperatures of $>$1700~K and COND \citep{allard2001,baraffe2003} models for $<$1700~K for translating magnitude limits into mass limits, given the estimated ages. The completeness of the survey as a whole can then be calculated as a 2D function of separation and mass by evaluating what fraction of the targets provide detectability for companions of any given mass and separation. We plot some contours of this function in Fig. \ref{f:completeness}. There are bumps in the individual contours due to strong PSF sidelobes and similar features in the images. As an example, we find that the completeness is 70\% for 10~$M_{\rm jup}$ planets outside of 1.2\arcsec, which corresponds to 160~AU at the median distance of the sample of 132~pc.

The detection limits quoted in terms of mass depend on the exact ages of the systems, hence given the uncertainty in age of the Sco-Cen association as discussed above, they should be treated with a certain degree of caution. However, the impact of the factor $\sim$2 uncertainty in age has a modest impact for the masses near our detection limits. For instance, according to the DUSTY models \citep{chabrier2000}, a 10~$M_{\rm jup}$ object at 5~Myr has the same K-band brightness as a $\sim$13~$M_{\rm jup}$ object at 10~Myr. Our 70\% completeness limit quoted above would thus only increase by $\sim$30\% if the \citet{pecaut2012} ages were applied.

\subsection{Statistical Properties}
\label{s:statistics}

\begin{figure}[p]
\centering
\includegraphics[width=8cm]{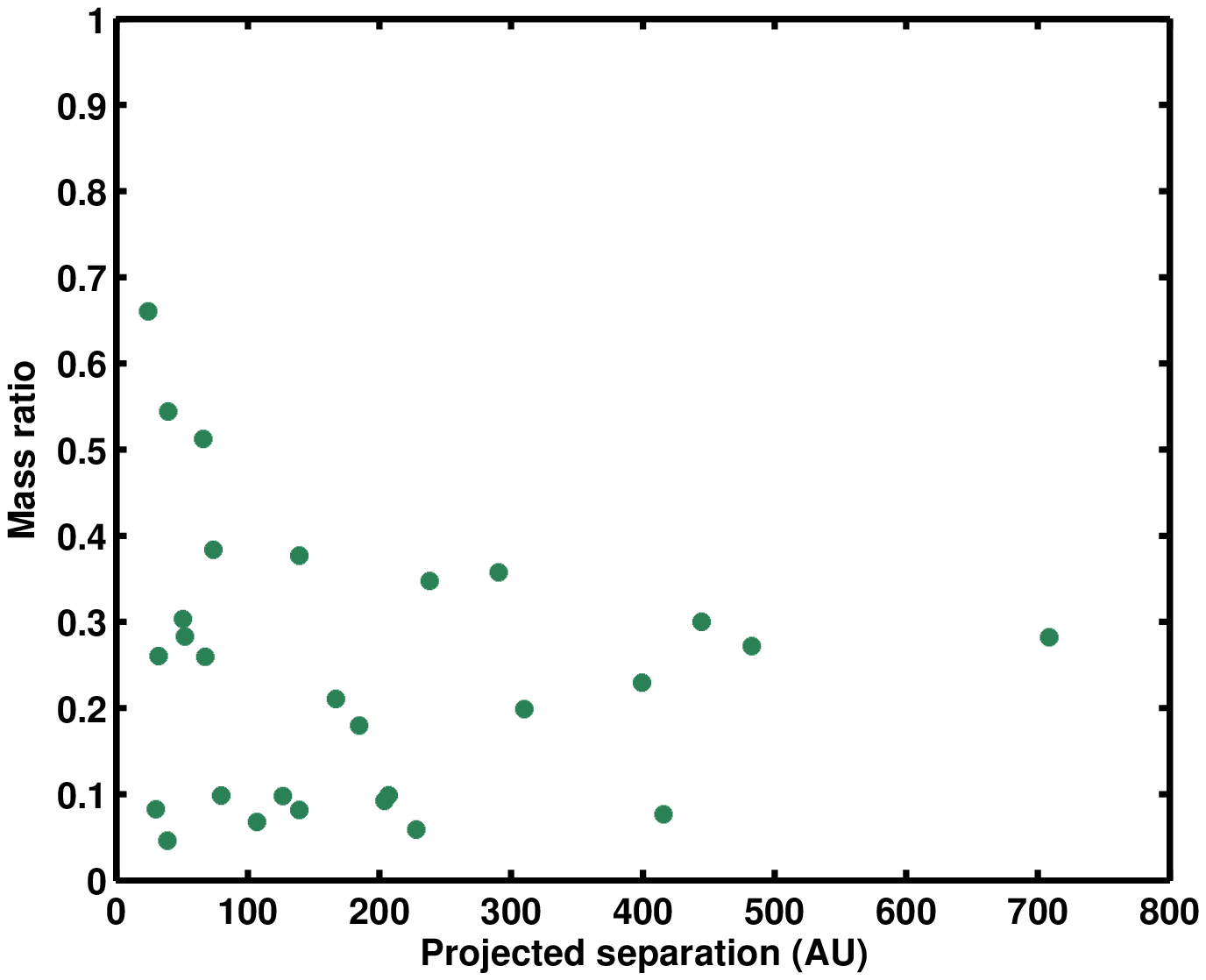}
\caption{Distributions in separation and mass ratio for the binaries in the sample. There is a lack of binaries with components of similar mass, particularly at large separations.}
\label{f:sepvsq}
\end{figure}

Here we will consider the statistical distributions of the binary population in our observed sample. Fig. \ref{f:sepvsq} displays the projected separation versus the mass fraction of the detected binaries. The most striking trend is a lack of nearly equal-mass binaries, particularly at large separations. This cannot be caused by any bias related to our detection limits, both because the completeness is very good over our whole considered separation range of 0.1\arcsec\ to 5.0\arcsec\ for masses down to our lowest-mass detected companions of $\sim$50~$M_{\rm jup}$ as shown in Sect. \ref{s:completeness}, and also because our sensitivity increases outward in separation and upward in mass, and is maximal in the range where the lack of companions is observed. The inferred masses of the individual components depend on the age, but since both components evolve with time, the impact of age uncertainties of the relevant order is limited on their calculated mass ratio. For instance, according to the \citet{baraffe1998} models, a 1~$M_{\rm sun}$ primary and 0.1~$M_{\rm sun}$ secondary at 10~Myr (the approximate UCL/LCC age according to \citet{song2012}) correspond in K-band brightness to a 1.1~$M_{\rm sun}$ primary and 0.13~$M_{\rm sun}$ secondary at 16~Myr (the corresponding \citet{pecaut2012} age). Hence, the mass ratio only changes from 0.10 to 0.12 between the younger and older age, and the impact decreases as the mass ratio gets larger, since the evolution of the components becomes more equal. As a result, the age uncertainty has a small impact on the observed mass ratio distribution, and cannot explain the lack of near equal-mass binaries.

We can attempt to quantify the companion mass ratio distribution with a simple power law, $f \sim q^{\alpha_q}$, where $f$ is the frequency, $q$ is the mass ratio (secondary to primary, i.e. between 0 and 1), and $\alpha_q$ the power law index. This is done by generating simulated populations with a given distribution determined by $\alpha_q$ and comparing to the observed population using a Kolmogorov-Smirnov test. For each population we generate 1000 binaries, and 1000 populations are being simulated in order to verify the robustness of the results. Cases where the secondary mass is $<$50~$M_{\rm jup}$ are removed in order to maintain a good completeness, hence the power law fit is only valid down to secondaries of this mass. The full range of separations (0.1--5.0\arcsec) is included in the comparison. The median probability among the 1000 simulations is adopted as the result of the full test, and the 5th and 95th percentiles are taken as the lower and upper bounds, respectively. This is equivalent to the procedure applied in \citet{babu2006}. In this way, we can firmly exclude that the mass ratio is uniformly distributed, since simulated populations with $\alpha_q = 0$ have a $1.4 \times 10^{-7} \pm 0.4 \times 10^{-7}$ probability of matching the observed distribution. It is necessary to make $\alpha_q$ substantially negative (bottom-heavy mass ratio distribution) to get better fits. A value of $\alpha_q = -0.4$ which has been used for more massive stars \citep[e.g.][]{kouwenhoven2007} is still too shallow, with a probability of 0.023$^{+0.007}_{-0.005}$\%. $\alpha_q = -0.9$ fits better at 17.5$^{+0.04}_{-0.03}$\%. However, this form of power law fitting does not have a promising asymptotic behaviour, as it starts to drastically overshoot at small mass ratios for steep power law indices. Hence, we slightly alter the form of the distribution, and formulate it as $f \sim 1 - q^{\beta_q}$, which better fits the observations for a range of values of $\beta_q$. Note that in this framework, a higher $\beta_q$ implies a more bottom-heavy mass ratio distribution, i.e. in the opposite sense to $\alpha_q$. Stepping through $\beta_q$ in steps of 0.5, we find that the best fit is provided by $\beta_q = 2.5$, which has a 66.6$^{+4.1}_{-3.8}$\% probability of matching the observed distribution, hence we adopt this value for the remainder of this analysis.

The semi-major axis distribution is another issue of interest to address. We do this in a very similar way as above, simulating populations with given distributions and comparing them to the observed distribution using a Kolmogorov-Smirnov test. For translating from our observed projected separations into semi-major axes, we multiply the separations by a factor 1.02. This was calculated based on \citet{brandeker2006} as the conversion factor between semi-major axis and projected separation for a sample with an assumed eccentricity distribution of $f(e) \sim 2e$. For a less eccentric distribution, the conversion factor can be somewhat higher, up to a maximum of 1.27 for completely circular orbits. For instance, \citet{fischer1992} calculate a factor 1.26 for their M-dwarf sample. We have tried the same analysis as below with a 1.26 conversion factor, and found that it makes very little difference for the results and has no impact on the conclusions. Hence, in the procedure described in the following, only the factor 1.02 is used. 

Relative to the mass ratio case, a comparison of the semi-major axis distribution to other distributions in the literature is less discriminating, partly due to the fact that the binaries here cover a somewhat limited semimajor axis range relative to the extremely wide range that occurs in the universe, spanning approximately six orders of magnitude from a few Solar radii to tens of thousands of AU. For Sun-like stars, a log-normal distribution with $\mu_a = 1.64$ and $\sigma_a = 1.52$ has been measured \citep{raghavan2010}. Applying this relation in our simulations results in a 29.9$^{+4.5}_{-3.9}$\% probability that it is drawn from the same distribution as our observed sample. For more massive stars, it has been suggested that {\"O}pik's law \citep{opik1924} might provide a better fit than a log-normal distribution \citep{kouwenhoven2007}. {\"O}pik's law represents a uniform distribution in logarithmic semi-major axis space. In this sense, it can be seen as a log-normal distribution with infinite $\sigma_a$. Applied to our sample, {\"O}pik's law results in a 39.3$^{+5.6}_{-5.4}$\% matching probability. Thus, while neither of these distributions fit extremely well, neither can be more than at best marginally excluded. This is consistent with a picture in which intermediate-mass objects have an intermediate distribution between lower-mass and higher-mass stars. We note that if we adopt a log-normal distribution and simply shift the Sun-like distribution to wider separations, $\mu_a = 2.40$ and $\sigma_a = 1.52$, we get a 67.2$\pm$5.1\% probability match, hence we adopt this relation for future purposes. In all of these cases, we set lower and upper bounds on the semi-major axis of 0.023~AU and 23000~AU, respectively \citep{kouwenhoven2007}.

The multiplicity fraction within 0.1\arcsec$< \rho <$5.0\arcsec\ is $27/130 = 20.8$\%, with 4.0\% errors assuming Poissonian statistics. If we were to assume a total multiplicity fraction of 100\%, then adopting the best-fit distributions in mass ratio and semi-major axis as above, and accounting for the completeness function, would result in a higher multiplicity fraction within 0.1\arcsec$< \rho <$5.0\arcsec\ of 35.1\%. Hence, these distributions imply that the actual total multiplicity should be approximately 59\%, in order to reproduce the 20.8\% multiplicity observed in our covered observational range. Obviously, there is significant uncertainty in this number, primarily due to the limited coverage in semi-major axis space. For instance, if we were to adopt {\"O}pik's law instead, then the total implied multiplicity fraction would be 81\%. However, since the distribution is as narrow as the Sun-like distribution in the former case, and it cannot possibly be broader than the {\"O}pik distribution adopted in the latter case, it is probably fair to conclude that the total multiplicity fraction is bounded between 59\% and 81\% for any realistic distribution.

\subsection{Individual Notes}
\label{s:notes}

Here we provide individual notes for targets for which special information exists.

\textbf{HIP~50083:} HIP~50083 is one of the four stars that have a mass of $>$5~$M_{\rm sun}$, and is therefore excluded from the statistical study. The image is of insufficient quality to put any stringent constraints on the presence or absence of companions.

\textbf{HIP~50847:} The image of HIP~50847 is of too poor quality to be scientifically useful. Nonetheless, there is a probable bright binary companion visible in the image, at a separation of $\sim$2.2\arcsec\ and a position angle of $\sim$351$^{\rm o}$. HIP~50847 is one of the four $>$5~$M_{\rm sun}$ stars that were omitted from any statistical analysis. Aside from the possible companion reported here, HIP~50847 (HD~90246) is a known double-lined spectroscopic binary with a period of $\sim$15~days \citep{quiroga2010}, hence the system is possibly triple.

\textbf{HIP~56543:} There is a bright source at the edge of the field, at a separation of 9.92\arcsec\ and a position angle of 310.7$^{\rm o}$. Thanks to the large separation and relatively small brightness contrast, this source is visible in 2MASS \citep{skrutskie2006} with designation 2MASS~J11353717-5043180, at a separation of 10.24\arcsec\ and position angle of 309.3$^{\rm o}$ from HIP~56543. The expected motion of HIP~56543 relative to a static background object over the $\sim$12 year baseline is 343~mas West and 31~mas South, fully in agreement with the observations. Hence, we can conclude that 2MASS~J11353717-5043180 is a background star, physically unrelated to HIP~56543.

\textbf{HIP~57238:} While the PSF of HIP~57238 is extended, it is extended to the same degree and in the same direction in both the primary and the well-resolved secondary in the system. Hence, we conclude that it is likely to be a PSF artefact.

\textbf{HIP~58899:} This is one of the two systems in the sample that are triple within the sensitivity range of the survey. For the statistical investigations of mass ratio and semi-major axis, HIP~58899 is counted as two pairs, where the tight AC pair is counted as a regular pair, and the wider AB pair is counted as another pair but the A+C mass is used for the A component.

\textbf{HIP~59481:} Given its relatively low contamination probability of 1.1\%, and the fact that it was the only point source in the field of view, the candidate companion to HIP~59481 was considered one of the most interesting targets for follow-up after the first epoch of imaging. However, the astrometric follow-up clearly shows that it is a background contaminant.

\textbf{HIP~63962:} Given a rather extended PSF, HIP~63962 may be an unresolved binary, with a separation of a few tens of mas (well under 50~mas) and a position angle of $\sim$40$^{\rm o}$.

\textbf{HIP~64322:} This star was observed under too poor conditions for the images to be scientifically useful, but a probable bright binary companion is visible at a separation of $\sim$2.3\arcsec\ and a position angle of $\sim$170$^{\rm o}$. Like all images of insufficient quality, it is omitted from the statistical analysis.

\textbf{HIP~66001:} Since the PSF of HIP~66001 is extended, it may be an unresolved binary. The separation would be a few tens of mas (well under 50~mas), and the position angle $\sim$30$^{\rm o}$.

\textbf{HIP~77038:} HIP~77038 is one out of the two systems in the survey that are triple systems. For statistical purposes, when counting mass ratio and semi-major axis, the system is counted as one pair in which the primary is one component, and the sum of the tight BC pair is the other component. The BC pair itself is not counted as a pair in this context, since the components are both late-type, and thus the mass of B is outside of the range of interest in primary mass.

\textbf{HIP~78384:} Also known as $\eta$~Lup, this is one of the four $>$5~$M_{\rm sun}$ stars that are excluded from our statistical study. The system is known as a probable triple system \citep[e.g.][]{eggleton2008}, but both the secondary and tertiary components are outside of the NICI field of view.

\textbf{HIP~79097:} Apart from the clearly resolved companion at 0.8\arcsec separation, there is an additional extension in the HIP~79097~A component, which is not present in the companion. This strongly implies that a third component is present in the system as a close companion to component A, with a separation of a few tens of mas (well under 50~mas) and a position angle of $\sim$110$^{\rm o}$.

\textbf{HIP~79977:} This star has a debris disk that was recently spatially resolved \citep{thalmann2013}. The deep image in \citet{thalmann2013} also has two faint point sources within 5\arcsec. Our image is too shallow to detect either the disk or the point sources at a statistically significant level, although the point sources can be traced at their expected background positions with prior knowledge of where to look.

\begin{figure}[p]
\centering
\includegraphics[width=8cm]{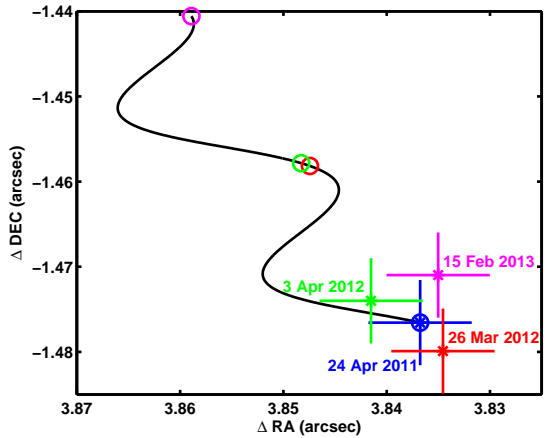}
\caption{Astrometry of the candidate companion to HIP~80130. The static background trajectory can be rejected by 10.0$\sigma$ in total.}
\label{f:astro80130}
\end{figure}

\begin{figure*}[p]
\centering
\includegraphics[width=8cm]{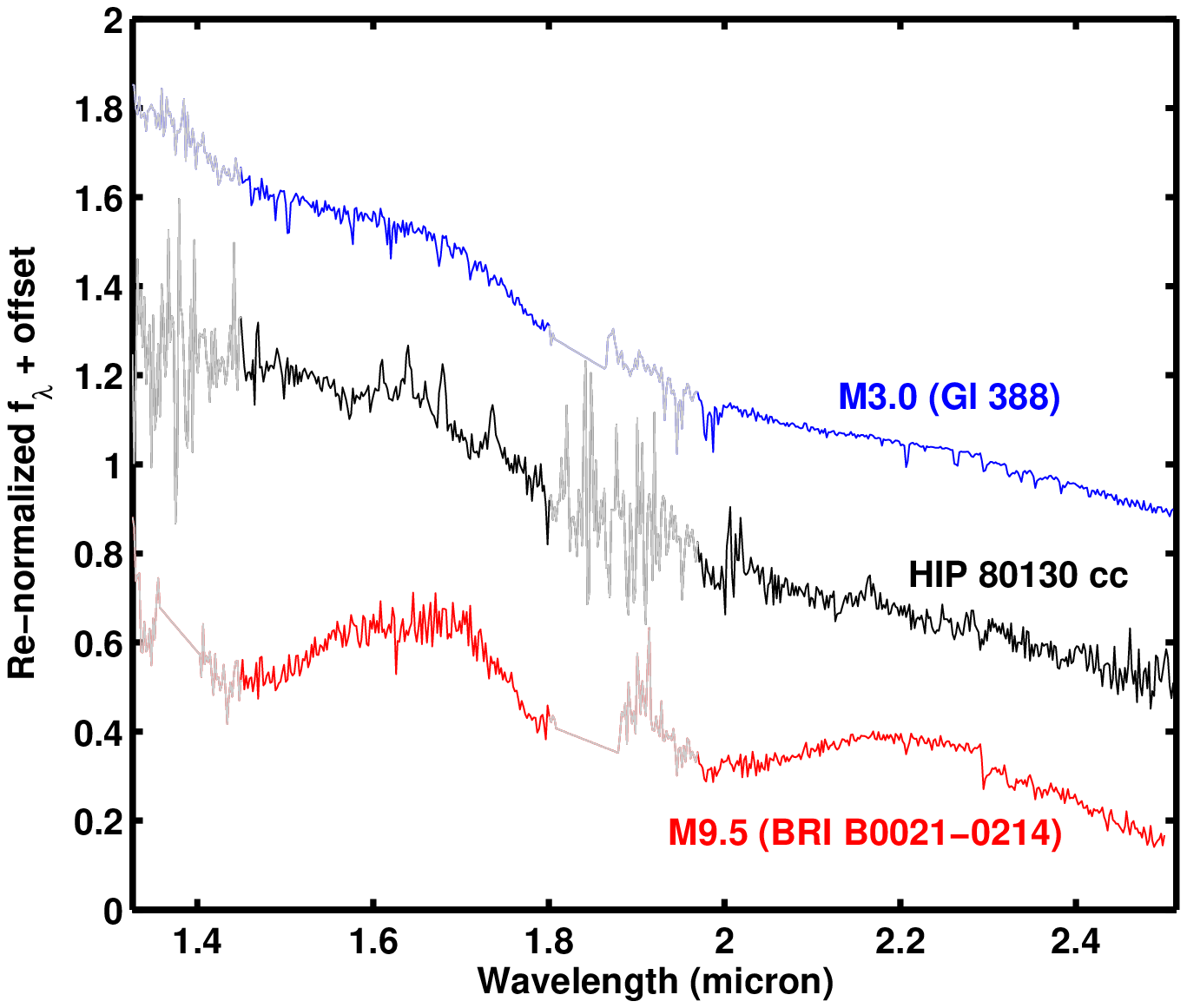}
\caption{Spectrum of the candidate companion to HIP~80130, along with spectral standard stars of spectral types M3V and M9.5V \citep{cushing2005,rayner2009}. The candidate spectrum fits best to an early M-type spectrum such as M3V, and is quite distinct from colder atmospheres such as M9.5V, even though the latter would be much more consistent with its $K$-band brightness.}
\label{f:spec80130}
\end{figure*}

\textbf{HIP~80130:} The most puzzling target in the survey, HIP~80130 has a faint point source residing at 4.1\arcsec separation from the star, which was identified as a candidate very low-mass brown dwarf in the first epoch image, and thus was followed up in a second epoch. The second epoch showed indications of common proper motion, so it was followed up in third (and subsequently a fourth) epoch of astrometry, and with spectroscopy. The photometry over the four epochs suggests strong variability, with $\Delta K = 7.1 \pm 0.8$~mag. The astrometry in all four epochs, spanning approximately two years, is essentially perfectly consistent with common proper motion, and is inconsistent with the expected background trajectory by a total of 10.0$\sigma$ (see Fig. \ref{f:astro80130}). On the other hand, the spectrum of the candidate favours a different interpretation. The best-fit spectral type is in the range of $\sim$M3 (see Fig. \ref{f:spec80130}), implying a temperature of 3300~K \citep{slesnick2004}. Using the \citet{baraffe1998} models and an age of 5~Myr, the implied mass is $\sim$250~$M_{\rm jup}$ and the corresponding predicted absolute magnitude is $M_{\rm K} = 5.25$~mag. This is wildly at conflict with the actual magnitude measured in the image (if we assume that the candidate is indeed comoving with HIP~80130) of $M_{\rm K} = 8.69$~mag, which by itself yields a mass of $\sim$18~$M_{\rm jup}$ from \citet{chabrier2000}. Thus, there is a discrepancy of 3.44~mag (a factor of 24) in brightness, or equivalently a factor of 14 in mass, between the estimations based on spectral type and apparent brightness, respectively. The gap is too large to be explained by any model uncertainties or errors in (e.g.) distance, or the variability of the source. 

A further observational constraint that underlines the peculiar nature of the source comes from the fact that HIP~80130 was covered in the UKIDSS survey \citep{lawrence2007} in a $K$-band image from epoch 2009.34. A point source is reported at a separation of 4.09\arcsec\ and a position angle of 111.0$^{\rm o}$, which corresponds to our identified point source. The location is considerably closer to the CPM location than the background trajectory, lending further support to the CPM hypothesis. The $K$-band contrast is 6.9~mag, consistent with our NICI contrast. Hence, the UKIDSS data broadly confirm the picture posed by the NICI data, that this is an unusual object in one way or another. The options that come most readily to mind are that the object is either a very unusual form of companion, or a background star which by chance has equal proper motion with HIP~80130, both of which seem highly unlikely. Perhaps the object is a 3300~K star which is obscured to an extreme degree (for its age), for instance by an edge-on disk. This might be supported by the strong variability. However, given these uncertainties, we cautiously do not include HIP~80130 in any of the statistical analysis for the moment, and we empasize that more data will be needed to clarify the real properties of the source.

\textbf{HIP~80208:} This is one of the four stars that have a mass of $>$5~$M_{\rm sun}$, and is therefore excluded from the statistical study.

\textbf{HIP~82569:} HIP~82569 resides in a crowded field, and has several point sources present within the NICI field of view. Surprisingly, none of the sources appear to move with respect to the primary over four epochs of imaging spanning nearly two years. While the proper motion of HIP~82569 is among the slowest in the sample at 24.4~mas/yr, it should be significant over the 2-year baseline. The normal inference for any single point source would be that it shares a common proper motion, and thus is a companion to the star. However, $\sim$10 point sources at similar projected separations, and some of which are evidently binaries themselves, cannot be companions to the same star. Hence, we consider the proper motion of this star to be unreliable and count the point sources as probable background sources. Further analysis in the future will be required to stringently test whether single sources in the field may nonetheless share a common proper motion with HIP~82569.

\section{Discussion}
\label{s:discussion}

The total multiplicity fraction that we derive of $\sim$60--80\% (depending on the semi-major axis distribution) fits very well with the general view that is emerging of a smoothly increasing multiplicity fraction with primary mass. This appears to hold from the brown dwarf range with derived multiplicities of $\sim$10--30\% \citep[e.g.][]{bouy2003,ahmic2007,joergens2008} through M-dwarfs with a multiplicity of $\sim$35\% \citep[e.g.][]{fischer1992,janson2012a} and a smooth increase from the lowest to highest M-dwarf masses \citep{janson2012a}, early K- through late F-stars with $\sim$40--50\% and again a smooth increase from the lower to the higher masses of the range \citep{raghavan2010}, our sample of F- and late A-type stars with 60--80\%, and to more massive A- and B-type stars with $\sim$80--100\% \citep[e.g.][]{shatsky2002,kouwenhoven2007} and once again an increase from lower to higher primary masses within the sample \citep{kouwenhoven2007}. The multiplicity of O-type stars is also consistently very high \citep[e.g. $>$80\% from SBs alone][]{chini2012}. Likewise, the semi-major axis distribution appears to fit in with the trend that the characteristic orbital size increases from brown dwarfs \citep{burgasser2007} through M-dwarfs \citep{janson2012a} and Sun-like stars \citep{raghavan2010}, since our data implies a further increase of orbital size around AF-type stars. For higher-mass stars, the distribution appears to be best described by {\"O}pik's law \citep{kouwenhoven2007}, which is uniform in logarithmic space rather than log-normal, hence the concept of a characteristic semi-major axis is not as well defined. Since our sample is marginally consistent with {\"O}pik's law, this again fits with the emerging picture that binary properties are universally continuous across all stellar and even brown dwarf properties.

However, one remaining puzzling issue with regards to our sample in particular is the mass ratio distribution. In principle, with the brown dwarf mass ratio distribution being strongly top-heavy and that of low-mass and Sun-like stars being essentially flat (\citealt{raghavan2010,janson2012a}; see also \citealt{goodwin2013,reggiani2013}), it would be natural to assume that AF-type stars should have a bottom-heavy distribution, as we do observe. However, the distribution is very steep, and significantly more so than what has been found for the higher-mass population \citep{kouwenhoven2007}, thus breaking the otherwise continuous trend. This effect appears to be physically real, since the most surprising feature of a lack of nearly equal-mass components occurs in the parameter space of maximal detectability. It also seems unlikely that it could be related to any selection bias, although there is, in fact, a selection issue related precisely to massive binary companions discussed in \citet{dezeeuw1999}, which they discuss for (e.g.) the case of USco member $\delta$~Sco. This system consists of a massive binary (primary spectral type is B0), unresolved to Hipparcos, where the orbital motion of the binary causes its photocenter motion to deviate from the motion of the center of mass over the relatively short timescales (3.3~years) covered by Hipparcos. In this case, the effect estimated by \citet{dezeeuw1999} is a 2~mas/yr deviation from the `real' proper motion, larger than the Hipparcos errors of $\sim$1~mas and sufficient to (presumably erroneously) discard it as a USco member based on their standard approach. Hence, other stars with massive companions may have been missed in their procedure, which would in turn potentially cause a selection bias against such targets in our sample. However, it is unlikely that it could be influencing this observational result. The stars in our sample are less massive than $\delta$~Sco by an order of magnitude, causing slower orbital motion for a given separation. Furthermore, the separations that we probe stretch out substantially larger than the 0.13\arcsec separation of $\delta$~Sco. As a result, we estimate that the magnitude of deviant motion for 1\arcsec\ separation binaries in our sample is an order of magnitude smaller than for $\delta$~Sco, which as mentioned above is 2~mas/yr. In this circumstance, they are much smaller than the astrometric errors of $\sim$1~mas/yr and cannot influence the kinematic classification of the primary star. Thus, while a selection bias could conceivably be present for small-separation binaries, it cannot affect the range of $>$1\arcsec\ binaries, where the lack of nearly equal-mass components is the most pronounced. Finally, we note that if such a bias were present, it should have affected the \citet{kouwenhoven2007} results in the same way, which appears not to be the case as \citet{kouwenhoven2007} derive a shallower mass ratio distribution. A possible solution can be seen in Fig. \ref{f:sepvsq}, where high mass ratios appear to become more common at smaller separations. The three cases of HIP~63962, HIP~66001, and HIP~79097, if interpreted as close binaries, would all have to have two nearly equal-mass components, which would further support this trend. Hence, it is possible that the mass ratio depends on semi-major axis and that our sample would yield a shallower distribution if observed over the full semi-major axis range rather than a quite limited range of it.

As a result of the detections of wide planetary mass objects and low-mass brown dwarfs such as 1RXS~J1609~b \citep{lafreniere2008a,lafreniere2010}, GSC 0621~B \citep{ireland2011}, and HIP~78530~B \citep{lafreniere2011}, it has been speculated that the Sco-Cen region might have a particularly high frequency of such companions, with \citet{ireland2011} citing a tentative frequency estimation of $\sim$4\% in the 200--500~AU range. If adopting this frequency, then given our high completeness to such companions and the fact that we survey 138 targets of which 130 are of sufficient quality, we should have expected to find $\sim$5 companions in the mass range between 1RXS~J1609~b and HIP~78530~B. Given this context, the fact that we find none is quite significant, with a $<$1\% probability of being consistent with a 4\% frequency. A few distinctions between the surveys are worthy of note. Firstly, the surveys in which the above detections were made were performed exclusively in the USco sub-region of Sco-Cen, whereas our survey was performed in the whole Sco-Cen region, with only a small minority of targets being specifically USco members. Thus, in principle there might be something special about USco, although this seems unlikely, since the sub-regions of Sco-Cen are very similar, must have formed under very similar conditions, and vary in age by only a factor of $\sim$2. Secondly, the stellar mass distributions are different between the surveys. However, while 1RXS~J1609~b and GSC~0621~B were discovered around Sun-like stars, HIP~78530~B was discovered around a B-type star, so the primary mass range at which detections have been made appear to encompass the range of AF-type stars covered in our survey. Hence, also for this distinction, it is unclear if it could explain the differences in frequency. Perhaps the most likely explanation is simply that the real frequency is intermediate, in the range of $\sim$2\%, and that modest statistical fluctuations caused the apparently dissimilar frequencies in the surveys.

\acknowledgements
We thank the staff at Gemini and ESO for their support during these observations. Our study made use of the CDS and SAO/NASA ADS services. Support for this work was provided by NASA through Hubble Fellowship grant HF-51290.01 awarded by the Space Telescope Science Institute, which is operated by the Association of Universities for Research in Astronomy, Inc., for NASA, under contract NAS 5-26555. Additional support came from grants to RJ from the Natural Sciences and Engineering Research Council of Canada.

\clearpage

\onecolumn
\begin{longtable}{lcccccccccc}
\caption{Target list.}\\
\hline\hline
Target & RA & Dec & SpT$^{\rm a}$ & Dist$^{\rm b}$ & $K$ & Assoc. & Age & Stat$^{\rm c}$ & Mult$^{\rm d}$ \\
 & (hh mm ss) & (dd mm ss) & & (pc) & (mag) & & (Myr) & & \\
\hline
\endfirsthead
\caption{continued.}\\
\hline\hline
Target & RA & Dec & SpT$^{\rm a}$ & Dist$^{\rm b}$ & $K$ & Assoc. & Age & Stat$^{\rm c}$ & Mult$^{\rm d}$ \\
 & (hh mm ss) & (dd mm ss) & & (pc) & (mag) & & (Myr) & & \\
\hline
\endhead
\hline
\endfoot
\hline
\multicolumn{10}{l}{\footnotesize{$^a$Spectral type from SIMBAD, complemented with values from \citet{chen2012}.}}\\
\multicolumn{10}{l}{\footnotesize{$^b$Distance uncertainty for individual stars is typically $\sim$20\%}}\\
\multicolumn{10}{l}{\footnotesize{$^c$Flag for whether the target is included in the statistical study, (Y)es or (N)o.}}\\
\multicolumn{10}{l}{\footnotesize{$^d$Flag for whether the target is observed to be multiple by NICI, (Y)es, (N)o, or (U)nclear.}}\\
\hline
\endlastfoot
HIP 50083	&	10 13 30.642	&	-66 22 22.12	&	A4	&	93	&	4.64	&	LCC	&	10	&	N	&	N	\\
HIP 50847	&	10 22 58.126	&	-66 54 05.31	&	B8	&	132	&	5.31	&	LCC	&	10	&	N	&	Y	\\
HIP 51991	&	10 37 20.537	&	-69 21 45.28	&	A1	&	177	&	8.01	&	LCC	&	10	&	Y	&	N	\\
HIP 55334	&	11 19 52.765	&	-70 37 06.54	&	F2	&	86	&	7.08	&	LCC	&	10	&	Y	&	N	\\
HIP 56543	&	11 35 38.011	&	-50 43 24.50	&	A5	&	137	&	7.77	&	LCC	&	10	&	Y	&	N	\\
HIP 57238	&	11 44 09.799	&	-53 44 54.42	&	A1	&	173	&	8.33	&	LCC	&	10	&	Y	&	Y	\\
HIP 57595	&	11 48 27.008	&	-54 09 18.52	&	F5	&	171	&	8.22	&	LCC	&	10	&	Y	&	Y	\\
HIP 57710	&	11 50 07.189	&	-49 32 35.55	&	A3	&	119	&	7.75	&	LCC	&	10	&	Y	&	Y	\\
HIP 57950	&	11 53 07.998	&	-56 43 38.14	&	F2	&	105	&	7.28	&	LCC	&	10	&	Y	&	N	\\
HIP 58075	&	11 54 35.595	&	-54 43 57.28	&	F2	&	159	&	7.93	&	LCC	&	10	&	Y	&	N	\\
HIP 58146	&	11 55 28.844	&	-62 11 47.19	&	F2	&	117	&	6.83	&	LCC	&	10	&	Y	&	N	\\
HIP 58167	&	11 55 43.547	&	-54 10 50.62	&	F3	&	91	&	7.28	&	LCC	&	10	&	Y	&	N	\\
HIP 58220	&	11 56 26.556	&	-58 49 16.87	&	F3	&	105	&	7.39	&	LCC	&	10	&	Y	&	Y	\\
HIP 58528	&	12 00 09.407	&	-57 07 02.18	&	F5	&	90	&	7.42	&	LCC	&	10	&	Y	&	Y	\\
HIP 58680	&	12 02 03.374	&	-51 49 15.71	&	A4	&	190	&	8.41	&	LCC	&	10	&	Y	&	N	\\
HIP 58899	&	12 04 44.470	&	-52 21 15.69	&	F3	&	114	&	7.35	&	LCC	&	10	&	Y	&	Y	\\
HIP 58996	&	12 05 47.483	&	-51 00 12.14	&	G2	&	102	&	7.31	&	LCC	&	10	&	Y	&	N	\\
HIP 59084	&	12 07 00.669	&	-59 41 40.80	&	F0	&	136	&	7.71	&	LCC	&	10	&	Y	&	N	\\
HIP 59481	&	12 11 58.825	&	-50 46 12.48	&	F3	&	117	&	7.51	&	LCC	&	10	&	Y	&	N	\\
HIP 59505	&	12 12 12.012	&	-54 13 49.49	&	A9	&	109	&	7.70	&	LCC	&	10	&	Y	&	N	\\
HIP 59693	&	12 14 28.644	&	-47 36 46.16	&	F6	&	119	&	8.30	&	LCC	&	10	&	Y	&	Y	\\
HIP 59716	&	12 14 50.713	&	-55 47 23.59	&	F5	&	101	&	7.28	&	LCC	&	10	&	Y	&	N	\\
HIP 59724	&	12 14 56.366	&	-47 56 54.59	&	A6	&	110	&	7.42	&	LCC	&	10	&	Y	&	N	\\
HIP 59960	&	12 17 53.190	&	-55 58 31.97	&	F5	&	92	&	6.68	&	LCC	&	10	&	Y	&	N	\\
HIP 60245	&	12 21 11.720	&	-48 03 19.24	&	F2	&	124	&	8.00	&	LCC	&	10	&	Y	&	N	\\
HIP 60348	&	12 22 24.847	&	-51 01 34.31	&	F5	&	78	&	7.67	&	LCC	&	10	&	Y	&	N	\\
HIP 60459	&	12 23 42.195	&	-63 52 12.16	&	A3	&	98	&	7.02	&	LCC	&	10	&	Y	&	N	\\
HIP 60513	&	12 24 18.290	&	-58 58 35.30	&	F3	&	126	&	7.46	&	LCC	&	10	&	Y	&	N	\\
HIP 60567	&	12 24 54.914	&	-52 00 15.76	&	F6	&	158	&	8.39	&	LCC	&	10	&	Y	&	N	\\
HIP 60885	&	12 28 40.057	&	-55 27 19.38	&	G1	&	142	&	7.29	&	LCC	&	10	&	Y	&	Y	\\
HIP 61049	&	12 30 46.269	&	-58 11 16.88	&	F7	&	100	&	7.07	&	LCC	&	10	&	Y	&	N	\\
HIP 61087	&	12 31 12.647	&	-61 54 31.55	&	F6	&	92	&	6.74	&	LCC	&	10	&	Y	&	N	\\
HIP 61684	&	12 38 42.793	&	-68 45 49.08	&	A9	&	106	&	7.20	&	LCC	&	10	&	Y	&	N	\\
HIP 62032	&	12 42 54.874	&	-50 49 00.07	&	F0	&	169	&	7.90	&	LCC	&	10	&	Y	&	Y	\\
HIP 62134	&	12 44 01.928	&	-53 30 20.53	&	F2	&	126	&	7.71	&	LCC	&	10	&	Y	&	N	\\
HIP 62171	&	12 44 26.593	&	-54 20 48.03	&	F3	&	114	&	7.75	&	LCC	&	10	&	Y	&	N	\\
HIP 62427	&	12 47 38.703	&	-58 24 56.74	&	F8	&	128	&	8.12	&	LCC	&	10	&	Y	&	N	\\
HIP 62657	&	12 50 19.719	&	-49 51 48.86	&	F5	&	106	&	7.72	&	LCC	&	10	&	Y	&	N	\\
HIP 62677	&	12 50 35.830	&	-68 05 28.85	&	F1	&	183	&	8.11	&	LCC	&	10	&	Y	&	N	\\
HIP 63041	&	12 55 03.916	&	-63 38 26.79	&	F0	&	96	&	7.14	&	LCC	&	10	&	Y	&	N	\\
HIP 63272	&	12 57 57.773	&	-52 36 54.65	&	F3	&	111	&	7.46	&	LCC	&	10	&	Y	&	Y	\\
HIP 63435	&	12 59 56.410	&	-50 54 35.05	&	F5	&	158	&	7.99	&	LCC	&	10	&	Y	&	N	\\
HIP 63439	&	12 59 59.876	&	-50 23 22.42	&	F4	&	135	&	8.04	&	LCC	&	10	&	Y	&	N	\\
HIP 63527	&	13 01 04.365	&	-53 08 08.48	&	F1	&	133	&	6.86	&	LCC	&	10	&	Y	&	N	\\
HIP 63836	&	13 04 59.449	&	-47 23 48.54	&	F7	&	106	&	7.87	&	LCC	&	10	&	Y	&	N	\\
HIP 63886	&	13 05 32.611	&	-58 32 07.87	&	F2	&	102	&	7.22	&	LCC	&	10	&	Y	&	N	\\
HIP 63962	&	13 06 27.408	&	-56 52 44.87	&	G0	&	211	&	7.80	&	LCC	&	10	&	Y	&	N	\\
HIP 64044	&	13 07 33.502	&	-52 54 19.85	&	F5	&	106	&	7.51	&	LCC	&	10	&	Y	&	N	\\
HIP 64184	&	13 09 16.200	&	-60 18 30.10	&	F3	&	83	&	7.16	&	LCC	&	10	&	Y	&	N	\\
HIP 64322	&	13 10 59.015	&	-62 05 15.77	&	F1	&	106	&	7.19	&	LCC	&	10	&	N	&	Y	\\
HIP 64877	&	13 17 55.420	&	-61 00 38.84	&	F5	&	114	&	7.41	&	LCC	&	10	&	Y	&	N	\\
HIP 64995	&	13 19 19.522	&	-59 28 20.24	&	F2	&	111	&	7.34	&	LCC	&	10	&	Y	&	N	\\
HIP 65136	&	13 20 51.617	&	-48 43 19.68	&	F0	&	162	&	8.33	&	LCC	&	10	&	Y	&	N	\\
HIP 65423	&	13 24 35.130	&	-55 57 24.24	&	G3	&	97	&	8.08	&	LCC	&	10	&	Y	&	Y	\\
HIP 65517	&	13 25 47.838	&	-48 14 57.74	&	G2	&	104	&	8.08	&	LCC	&	10	&	Y	&	Y	\\
HIP 65617	&	13 27 12.195	&	-59 38 14.23	&	F8	&	164	&	8.36	&	LCC	&	10	&	Y	&	N	\\
HIP 65875	&	13 30 08.977	&	-58 29 04.34	&	F6	&	97	&	6.90	&	LCC	&	10	&	Y	&	N	\\
HIP 66001	&	13 31 53.609	&	-51 13 33.06	&	G8	&	167	&	7.83	&	LCC	&	10	&	Y	&	N	\\
HIP 67230	&	13 46 35.397	&	-62 04 09.64	&	F5	&	135	&	6.89	&	LCC	&	10	&	Y	&	N	\\
HIP 67428	&	13 49 09.222	&	-54 13 42.30	&	F5	&	117	&	7.63	&	LCC	&	10	&	Y	&	Y	\\
HIP 67497	&	13 49 54.502	&	-50 14 23.83	&	F0	&	106	&	7.52	&	UCL	&	10	&	Y	&	N	\\
HIP 67970	&	13 55 09.996	&	-50 44 42.94	&	F3	&	111	&	7.68	&	UCL	&	10	&	Y	&	N	\\
HIP 68722	&	14 04 04.926	&	-37 17 00.78	&	A7	&	149	&	7.72	&	UCL	&	10	&	Y	&	N	\\
HIP 69291	&	14 10 59.612	&	-36 16 01.68	&	F2	&	140	&	7.67	&	UCL	&	10	&	Y	&	Y	\\
HIP 69302	&	14 11 04.886	&	-49 16 23.40	&	A8	&	148	&	7.71	&	UCL	&	10	&	N	&	N	\\
HIP 69327	&	14 11 19.987	&	-54 37 56.06	&	F0	&	135	&	7.71	&	UCL	&	10	&	N	&	N	\\
HIP 69605	&	14 14 45.490	&	-38 22 52.37	&	A9	&	200	&	7.94	&	UCL	&	10	&	Y	&	N	\\
HIP 69720	&	14 16 16.986	&	-53 49 02.15	&	F0	&	140	&	7.86	&	UCL	&	10	&	Y	&	N	\\
HIP 70149	&	14 21 11.537	&	-41 42 24.96	&	A9	&	115	&	8.11	&	UCL	&	10	&	Y	&	N	\\
HIP 70558	&	14 25 58.517	&	-44 49 23.22	&	F2	&	122	&	8.13	&	UCL	&	10	&	Y	&	N	\\
HIP 71453	&	14 36 44.131	&	-40 12 41.61	&	B8	&	129	&	6.02	&	UCL	&	10	&	Y	&	N	\\
HIP 71498	&	14 37 19.427	&	-54 53 50.19	&	A2	&	155	&	8.22	&	UCL	&	10	&	Y	&	N	\\
HIP 71708	&	14 40 05.026	&	-40 54 02.30	&	A5	&	129	&	7.79	&	UCL	&	10	&	Y	&	Y	\\
HIP 71767	&	14 40 45.933	&	-42 47 06.32	&	F3	&	193	&	7.77	&	UCL	&	10	&	Y	&	Y	\\
HIP 72099	&	14 44 56.874	&	-34 22 53.76	&	F6	&	157	&	8.40	&	UCL	&	10	&	Y	&	Y	\\
HIP 72584	&	14 50 30.562	&	-35 05 36.30	&	A2	&	173	&	7.60	&	UCL	&	10	&	Y	&	N	\\
HIP 72630	&	14 51 00.664	&	-36 23 06.50	&	A9	&	165	&	8.45	&	UCL	&	10	&	Y	&	N	\\
HIP 73147	&	14 56 55.776	&	-42 27 40.38	&	A1	&	217	&	7.94	&	UCL	&	10	&	Y	&	N	\\
HIP 73777	&	15 04 48.919	&	-39 49 23.59	&	F8	&	94	&	8.16	&	UCL	&	10	&	Y	&	N	\\
HIP 73913	&	15 06 17.952	&	-35 24 22.27	&	A9	&	141	&	7.83	&	UCL	&	10	&	Y	&	N	\\
HIP 73990	&	15 07 14.943	&	-29 30 16.07	&	A9	&	97	&	7.32	&	UCL	&	10	&	Y	&	N	\\
HIP 74104	&	15 08 42.506	&	-44 29 04.49	&	A2	&	168	&	7.67	&	UCL	&	10	&	Y	&	Y	\\
HIP 74499	&	15 13 27.961	&	-33 08 50.23	&	F4	&	114	&	7.65	&	UCL	&	10	&	Y	&	N	\\
HIP 74865	&	15 17 56.113	&	-30 28 41.49	&	F3	&	124	&	7.81	&	UCL	&	10	&	Y	&	N	\\
HIP 74959	&	15 19 05.423	&	-36 21 44.08	&	F5	&	151	&	8.15	&	UCL	&	10	&	Y	&	N	\\
HIP 75480	&	15 25 09.398	&	-26 34 31.05	&	F0	&	136	&	7.38	&	UCL	&	10	&	Y	&	N	\\
HIP 75491	&	15 25 16.053	&	-38 09 28.64	&	F3	&	173	&	7.43	&	UCL	&	10	&	Y	&	N	\\
HIP 75683	&	15 27 42.320	&	-36 14 13.12	&	F4	&	141	&	8.37	&	UCL	&	10	&	Y	&	N	\\
HIP 75824	&	15 29 23.098	&	-40 09 49.96	&	F3	&	152	&	7.77	&	UCL	&	10	&	Y	&	N	\\
HIP 75891	&	15 30 04.278	&	-41 07 10.16	&	F2	&	155	&	7.57	&	UCL	&	10	&	Y	&	Y	\\
HIP 75933	&	15 30 34.047	&	-38 29 46.32	&	F3	&	194	&	7.72	&	UCL	&	10	&	Y	&	N	\\
HIP 76084	&	15 32 20.139	&	-31 08 33.75	&	F2	&	146	&	7.51	&	UCL	&	10	&	Y	&	N	\\
HIP 76501	&	15 37 27.916	&	-32 29 06.10	&	F2	&	154	&	7.52	&	UCL	&	10	&	Y	&	N	\\
HIP 76875	&	15 41 53.217	&	-34 53 19.91	&	F2	&	97	&	7.41	&	UCL	&	10	&	Y	&	N	\\
HIP 77038	&	15 43 47.639	&	-35 28 29.88	&	F3	&	128	&	7.92	&	UCL	&	10	&	Y	&	Y	\\
HIP 77388	&	15 47 51.175	&	-38 15 36.09	&	A6	&	136	&	7.40	&	UCL	&	10	&	Y	&	Y	\\
HIP 77432	&	15 48 24.788	&	-42 37 04.97	&	F5	&	99	&	7.87	&	UCL	&	10	&	Y	&	N	\\
HIP 77502	&	15 49 31.985	&	-31 15 39.69	&	F3	&	200	&	7.87	&	UCL	&	10	&	Y	&	N	\\
HIP 77520	&	15 49 39.636	&	-38 46 39.15	&	F3	&	131	&	8.00	&	UCL	&	10	&	Y	&	Y	\\
HIP 77545	&	15 49 59.797	&	-25 09 03.39	&	A2	&	111	&	7.90	&	US	&	5	&	Y	&	N	\\
HIP 77713	&	15 51 59.758	&	-34 49 41.44	&	F5	&	136	&	8.11	&	UCL	&	10	&	Y	&	N	\\
HIP 77813	&	15 53 20.898	&	-19 23 53.58	&	F8	&	90	&	7.30	&	US	&	5	&	Y	&	N	\\
HIP 77815	&	15 53 21.926	&	-21 58 16.54	&	A5	&	143	&	7.22	&	US	&	5	&	Y	&	Y	\\
HIP 78043	&	15 56 05.616	&	-36 53 34.53	&	F3	&	172	&	7.94	&	UCL	&	10	&	Y	&	N	\\
HIP 78150	&	15 57 28.548	&	-50 16 10.66	&	A7	&	146	&	7.21	&	UCL	&	10	&	Y	&	Y	\\
HIP 78233	&	15 58 29.305	&	-21 24 03.97	&	F2	&	258	&	7.69	&	US	&	5	&	Y	&	N	\\
HIP 78324	&	15 59 30.880	&	-40 51 54.57	&	B9	&	167	&	7.49	&	UCL	&	10	&	Y	&	N	\\
HIP 78384	&	16 00 07.322	&	-38 23 48.04	&	B2	&	151	&	4.09	&	UCL	&	10	&	N	&	N	\\
HIP 78555	&	16 02 18.532	&	-35 16 11.74	&	F0	&	101	&	7.73	&	UCL	&	10	&	Y	&	N	\\
HIP 78641	&	16 03 13.550	&	-35 17 14.90	&	A5	&	132	&	7.62	&	UCL	&	10	&	Y	&	N	\\
HIP 78963	&	16 07 12.673	&	-27 05 58.02	&	A9	&	169	&	7.31	&	US	&	5	&	Y	&	N	\\
HIP 78977	&	16 07 17.788	&	-22 03 36.49	&	F7	&	114	&	7.05	&	US	&	5	&	Y	&	N	\\
HIP 79054	&	16 08 10.509	&	-23 51 02.44	&	F0	&	126	&	7.83	&	US	&	5	&	Y	&	Y	\\
HIP 79083	&	16 08 35.144	&	-20 45 29.64	&	F4	&	159	&	6.68	&	US	&	5	&	Y	&	N	\\
HIP 79097	&	16 08 43.660	&	-25 22 36.74	&	F3	&	171	&	7.25	&	US	&	5	&	Y	&	Y	\\
HIP 79258	&	16 10 35.957	&	-32 45 42.78	&	F3	&	144	&	8.21	&	US	&	5	&	Y	&	N	\\
HIP 79288	&	16 10 55.110	&	-25 31 21.41	&	F0	&	116	&	7.88	&	US	&	5	&	Y	&	N	\\
HIP 79369	&	16 11 55.519	&	-21 06 17.99	&	F0	&	122	&	7.56	&	US	&	5	&	Y	&	N	\\
HIP 79392	&	16 12 09.894	&	-23 55 17.54	&	A2	&	194	&	7.54	&	US	&	5	&	Y	&	Y	\\
HIP 79516	&	16 13 34.331	&	-45 49 03.59	&	F5	&	127	&	7.79	&	UCL	&	10	&	Y	&	N	\\
HIP 79606	&	16 14 40.161	&	-20 14 03.00	&	F6	&	179	&	7.07	&	US	&	5	&	Y	&	N	\\
HIP 79673	&	16 15 37.144	&	-41 38 58.56	&	F2	&	146	&	7.84	&	UCL	&	10	&	Y	&	N	\\
HIP 79710	&	16 16 03.841	&	-49 04 29.38	&	F0	&	103	&	7.61	&	UCL	&	10	&	Y	&	N	\\
HIP 79733	&	16 16 21.918	&	-28 09 50.48	&	A0	&	249	&	7.93	&	US	&	5	&	Y	&	N	\\
HIP 79742	&	16 16 28.381	&	-38 44 12.32	&	F5	&	141	&	8.07	&	UCL	&	10	&	Y	&	N	\\
HIP 79910	&	16 18 39.147	&	-21 35 34.18	&	F3	&	138	&	7.54	&	US	&	5	&	Y	&	N	\\
HIP 79977	&	16 19 29.237	&	-21 24 13.25	&	F2	&	132	&	7.80	&	US	&	5	&	Y	&	N	\\
HIP 80088	&	16 20 50.226	&	-22 35 38.73	&	A9	&	164	&	7.79	&	US	&	5	&	Y	&	N	\\
HIP 80130	&	16 21 21.148	&	-22 06 32.26	&	A9	&	136	&	7.37	&	US	&	5	&	N	&	U	\\
HIP 80208	&	16 22 27.994	&	-49 34 20.51	&	B6	&	171	&	5.40	&	UCL	&	10	&	N	&	N	\\
HIP 80586	&	16 27 12.528	&	-27 11 21.96	&	F5	&	128	&	7.19	&	US	&	5	&	Y	&	N	\\
HIP 81392	&	16 37 21.670	&	-30 06 52.19	&	G2	&	197	&	7.78	&	US	&	5	&	Y	&	N	\\
HIP 81455	&	16 38 10.816	&	-29 40 40.20	&	F5	&	115	&	8.04	&	US	&	5	&	Y	&	N	\\
HIP 81851	&	16 43 05.389	&	-26 27 30.80	&	F2	&	126	&	7.51	&	US	&	5	&	Y	&	N	\\
HIP 82218	&	16 47 47.338	&	-19 52 31.98	&	F2	&	127	&	7.80	&	US	&	5	&	Y	&	N	\\
HIP 82534	&	16 52 13.311	&	-26 55 10.86	&	F0	&	150	&	7.37	&	US	&	5	&	Y	&	N	\\
HIP 82569	&	16 52 41.719	&	-38 45 37.30	&	F3	&	152	&	7.56	&	UCL	&	10	&	Y	&	N	\\
HIP 83159	&	16 59 42.481	&	-37 26 16.88	&	F5	&	148	&	7.92	&	UCL	&	10	&	Y	&	N	\\
\label{t:sample}
\end{longtable}
\twocolumn

\onecolumn
\begin{longtable}{lccccccccc}
\caption{Candidate astrometry}\\
\hline\hline
Target & CC & $\Delta K^{\rm a}$ & Epoch 1 & Sep$^{\rm a}$ & PA$^{\rm a}$ & Epoch 2 & Sep$^{\rm a}$ & PA$^{\rm a}$ & FAP \\
 & & (mag) & (MJD) & (\arcsec) & (deg) & (MJD) & (\arcsec) & (deg) & (\%) \\
\hline
\endfirsthead
\caption{continued.}\\
\hline\hline
Target & CC & $\Delta K^{\rm a}$ & Epoch 1 & Sep$^{\rm a}$ & PA$^{\rm a}$ & Epoch 2 & Sep$^{\rm a}$ & PA$^{\rm a}$ & FAP$^{\rm b}$ \\
 & & (mag) & (MJD) & (\arcsec) & (deg) & (MJD) & (\arcsec) & (deg) & (\%) \\
\hline
\endhead
\hline
\endfoot
\hline
\multicolumn{8}{l}{\footnotesize{$^a$Errors are 0.2~mag in contrast, 13~mas in separation and 0.4--0.7$^{\rm o}$ in position angle.}}\\
\multicolumn{8}{l}{\footnotesize{$^b$False Alarm Probability, see text for details.}}\\
\multicolumn{8}{l}{\footnotesize{$^c$Strongly variable, see individual note.}}\\
\hline
\endlastfoot
HIP 55334	&	1	&	9.5	&	55641.0	&	3.346	&	42.2	&	56053.0	&	3.347	&	43.0	&	9.8	\\
HIP 56543	&	1	&	9.0	&	55641.0	&	2.736	&	239.1	&	55987.2	&	2.722	&	238.9	&	5.4	\\
HIP 56543	&	2	&	9.4	&	55641.0	&	2.684	&	242.1	&	55987.2	&	2.668	&	241.8	&	7.0	\\
HIP 57595	&	1	&	7.9	&	55663.0	&	3.345	&	84.5	&	55987.2	&	3.373	&	84.5	&	7.0	\\
HIP 57595	&	2	&	8.1	&	55663.0	&	4.966	&	314.7	&	55987.2	&	4.979	&	315.0	&	17.0	\\
HIP 58220	&	1	&	3.6	&	55641.2	&	0.761	&	315.9	&	55987.5	&	0.751	&	316.3	&	0.02	\\
HIP 58220	&	2	&	10.1	&	55641.2	&	2.752	&	16.7	&	55987.5	&	2.748	&	17.4	&	33.1	\\
HIP 58220	&	3	&	11.8	&	55641.2	&	2.887	&	164.1	&	55987.5	&	2.911	&	163.1	&	80.8	\\
HIP 58220	&	4	&	11.4	&	55641.2	&	4.805	&	197.6	&	55987.5	&	4.802	&	197.0	&	96.5	\\
HIP 58680	&	1	&	8.3	&	55641.2	&	4.433	&	149.8	&	56020.3	&	4.456	&	149.7	&	13.6	\\
HIP 59481	&	1	&	8.0	&	55641.2	&	2.213	&	174.1	&	55990.2	&	2.226	&	173.3	&	1.1	\\
HIP 59716	&	1	&	11.0	&	55641.2	&	2.409	&	160.3	&	55990.2	&	2.428	&	159.5	&	23.4	\\
HIP 60245	&	1	&	11.1	&	55641.2	&	2.027	&	193.6	&	56020.2	&	2.023	&	192.6	&	16.2	\\
HIP 60348	&	1	&	10.5	&	55641.2	&	3.809	&	172.9	&	56015.2	&	3.819	&	172.4	&	25.5	\\
HIP 60459	&	1	&	9.0	&	55650.0	&	1.606	&	75.6	&	55990.2	&	1.616	&	75.3	&	13.6	\\
HIP 60459	&	2	&	9.6	&	55650.0	&	2.273	&	225.3	&	55990.2	&	2.250	&	225.6	&	38.2	\\
HIP 60459	&	3	&	8.5	&	55650.0	&	4.547	&	13.4	&	55990.2	&	4.563	&	13.5	&	54.2	\\
HIP 60513	&	1	&	9.1	&	55663.0	&	1.927	&	284.6	&	55990.2	&	1.911	&	285.6	&	9.1	\\
HIP 61087	&	1	&	6.8	&	55663.0	&	1.284	&	149.4	&	55990.2	&	1.249	&	148.1	&	1.1	\\
HIP 61087	&	2	&	8.5	&	55663.0	&	2.383	&	101.7	&	55990.2	&	2.387	&	100.6	&	13.1	\\
HIP 61087	&	3	&	7.5	&	55663.0	&	4.269	&	107.5	&	55990.2	&	4.273	&	107.0	&	18.5	\\
HIP 61087	&	4	&	9.3	&	55663.0	&	2.724	&	83.3	&	55990.2	&	2.744	&	82.6	&	29.2	\\
HIP 61087	&	5	&	9.8	&	55663.0	&	4.890	&	83.2	&	55990.2	&	4.902	&	82.8	&	80.9	\\
HIP 62171	&	1	&	9.8	&	55663.0	&	2.968	&	289.7	&	56015.2	&	2.969	&	290.3	&	14.9	\\
HIP 62677	&	1	&	7.2	&	55663.0	&	1.628	&	147.9	&	56020.2	&	1.645	&	147.3	&	2.1	\\
HIP 62677	&	2	&	5.8	&	55663.0	&	3.317	&	154.4	&	56020.2	&	3.333	&	154.5	&	3.1	\\
HIP 62677	&	3	&	7.3	&	55663.0	&	3.470	&	220.9	&	56020.2	&	3.452	&	220.6	&	10.0	\\
HIP 63041	&	1	&	9.4	&	55685.0	&	1.859	&	211.4	&	56020.2	&	1.844	&	211.4	&	32.7	\\
HIP 63041	&	2	&	8.5	&	55685.0	&	2.468	&	192.4	&	56020.2	&	2.458	&	192.1	&	28.9	\\
HIP 63041	&	3	&	9.0	&	55685.0	&	4.130	&	136.6	&	56020.2	&	4.144	&	136.4	&	75.9	\\
HIP 63041	&	4	&	7.8	&	55685.0	&	2.830	&	48.3	&	56020.2	&	2.843	&	49.5	&	22.7	\\
HIP 63836	&	1	&	7.2	&	55695.2	&	4.964	&	36.4	&	55987.2	&	4.995	&	36.5	&	3.5	\\
HIP 63886	&	1	&	6.2	&	55695.2	&	4.966	&	29.2	&	56053.0	&	5.007	&	29.4	&	5.5	\\
HIP 63886	&	2	&	9.3	&	55695.2	&	4.842	&	325.0	&	56053.0	&	4.853	&	325.1	&	43.4	\\
HIP 64184	&	1	&	8.6	&	55725.0	&	1.480	&	155.7	&	55987.2	&	1.472	&	155.5	&	5.0	\\
HIP 64184	&	2	&	9.6	&	55725.0	&	2.050	&	345.8	&	55987.2	&	2.075	&	347.1	&	18.9	\\
HIP 64877	&	1	&	9.8	&	55654.2	&	2.802	&	288.6	&	55987.2	&	2.768	&	288.9	&	57.7	\\
HIP 64995	&	1	&	6.8	&	55654.2	&	3.264	&	63.9	&	55987.2	&	3.280	&	63.7	&	6.0	\\
HIP 65423	&	1	&	4.4	&	55654.2	&	1.832	&	247.3	&	55987.2	&	1.832	&	247.6	&	0.2	\\
HIP 65517	&	1	&	4.5	&	55725.0	&	0.352	&	321.9	&	55987.2	&	0.354	&	322.5	&	0.004	\\
HIP 65617	&	1	&	8.3	&	56024.0	&	1.237	&	319.9	&	56336.5	&	1.235	&	320.5	&	6.9	\\
HIP 65617	&	2	&	9.3	&	56024.0	&	1.545	&	53.7	&	56336.5	&	1.570	&	53.7	&	21.5	\\
HIP 65617	&	3	&	8.4	&	56024.0	&	1.139	&	134.0	&	56336.5	&	1.147	&	133.8	&	6.4	\\
HIP 67230	&	1	&	9.7	&	56024.0	&	2.866	&	158.6	&	56336.5	&	2.880	&	158.1	&	59.6	\\
HIP 67230	&	2	&	9.5	&	56024.0	&	2.964	&	126.0	&	56336.5	&	2.989	&	125.7	&	56.1	\\
HIP 69605	&	1	&	10.0	&	56022.0	&	2.043	&	105.1	&	56339.5	&	2.065	&	104.9	&	4.0	\\
HIP 69720	&	1	&	10.9	&	56024.2	&	4.214	&	345.9	&	56336.5	&	4.230	&	346.2	&	75.1	\\
HIP 69720	&	2	&	11.9	&	56024.2	&	4.690	&	308.2	&	56336.5	&	4.683	&	308.6	&	97.5	\\
HIP 70149	&	1	&	9.8	&	56022.0	&	2.136	&	185.4	&	56339.5	&	2.110	&	184.9	&	5.2	\\
HIP 71498	&	1	&	8.8	&	56024.2	&	3.875	&	129.3	&	56339.5	&	3.870	&	128.9	&	41.0	\\
HIP 71498	&	2	&	10.5	&	56024.2	&	4.438	&	294.6	&	56339.5	&	4.433	&	294.8	&	91.6	\\
HIP 72099	&	1	&	4.2	&	55723.2	&	0.664	&	33.8	&	55987.2	&	0.671	&	34.4	&	0.007	\\
HIP 72584	&	1	&	10.3	&	55723.2	&	2.612	&	61.7	&	56020.2	&	2.635	&	61.7	&	5.7	\\
HIP 72630	&	1	&	4.3	&	55677.2	&	4.350	&	176.3	&	55987.2	&	4.404	&	176.2	&	0.3	\\
HIP 72630	&	2	&	9.7	&	55677.2	&	4.626	&	184.8	&	55987.2	&	4.662	&	184.5	&	30.1	\\
HIP 73147	&	1	&	7.5	&	55677.2	&	4.143	&	254.2	&	55987.2	&	4.098	&	254.5	&	4.8	\\
HIP 73147	&	2	&	10.2	&	55677.2	&	4.200	&	306.0	&	55987.2	&	4.194	&	306.5	&	34.2	\\
HIP 73777	&	1	&	8.6	&	55677.2	&	3.267	&	24.3	&	55987.2	&	3.312	&	24.3	&	5.8	\\
HIP 73913	&	1	&	10.1	&	55677.2	&	3.645	&	108.5	&	56013.2	&	3.672	&	108.3	&	17.8	\\
HIP 74959	&	1	&	10.3	&	55715.2	&	3.095	&	43.7	&	56020.2	&	3.127	&	43.3	&	18.7	\\
HIP 74959	&	2	&	10.1	&	55715.2	&	3.370	&	229.2	&	56020.2	&	3.338	&	229.7	&	18.9	\\
HIP 75683	&	1	&	9.2	&	55715.2	&	2.209	&	69.8	&	56005.2	&	2.211	&	69.3	&	5.8	\\
HIP 75683	&	2	&	8.9	&	55715.2	&	4.232	&	261.8	&	56005.2	&	4.226	&	262.2	&	15.9	\\
HIP 75824	&	1	&	7.5	&	55723.2	&	3.016	&	36.5	&	56015.2	&	3.022	&	36.9	&	2.5	\\
HIP 75824	&	2	&	11.1	&	55723.2	&	4.333	&	63.4	&	56015.2	&	4.338	&	63.7	&	56.3	\\
HIP 75891	&	1	&	2.5	&	55723.2	&	0.438	&	314.0	&	55990.2	&	0.450	&	312.7	&	0.001	\\
HIP 75933	&	2	&	10.5	&	55723.2	&	3.178	&	2.6	&	56012.2	&	3.195	&	3.5	&	18.5	\\
HIP 75933	&	3	&	12.0	&	55723.2	&	4.406	&	88.5	&	56012.2	&	4.433	&	88.9	&	69.8	\\
HIP 76501	&	1	&	11.1	&	55723.2	&	3.229	&	131.8	&	56005.2	&	3.229	&	132.1	&	16.8	\\
HIP 76501	&	2	&	11.3	&	55723.2	&	4.587	&	279.6	&	56005.2	&	4.582	&	280.2	&	34.8	\\
HIP 77038	&	1	&	3.6	&	56024.2	&	1.443	&	232.3	&	56335.2	&	1.437	&	232.5	&	0.02	\\
HIP 77038	&	2	&	4.5	&	56024.2	&	1.429	&	243.2	&	56335.2	&	1.408	&	243.8	&	0.05	\\
HIP 77432	&	1	&	8.3	&	55712.2	&	3.454	&	159.0	&	56005.2	&	3.427	&	159.1	&	11.0	\\
HIP 77432	&	2	&	8.6	&	55712.2	&	1.764	&	265.8	&	56005.2	&	1.777	&	266.7	&	3.7	\\
HIP 77432	&	3	&	9.5	&	55712.2	&	4.264	&	69.7	&	56005.2	&	4.259	&	69.5	&	35.7	\\
HIP 77502	&	1	&	9.0	&	56024.2	&	1.213	&	59.9	&	56338.5	&	1.224	&	59.6	&	0.8	\\
HIP 77520	&	1	&	1.9	&	55712.2	&	2.222	&	196.7	&	56053.0	&	2.225	&	196.7	&	0.03	\\
HIP 77520	&	2	&	9.7	&	55712.2	&	3.535	&	174.9	&	56053.0	&	3.537	&	174.4	&	21.4	\\
HIP 77713	&	1	&	9.4	&	55712.2	&	2.949	&	62.0	&	56005.2	&	2.972	&	61.8	&	10.2	\\
HIP 78324	&	1	&	10.1	&	55712.2	&	3.492	&	110.9	&	56005.2	&	3.492	&	111.1	&	29.7	\\
HIP 78555	&	1	&	7.6	&	55712.2	&	1.954	&	71.6	&	56005.2	&	1.965	&	70.8	&	1.2	\\
HIP 78555	&	2	&	7.7	&	55712.2	&	4.399	&	185.1	&	56005.2	&	4.371	&	185.1	&	6.2	\\
HIP 78555	&	3	&	10.3	&	55712.2	&	3.740	&	281.9	&	56005.2	&	3.751	&	282.6	&	29.3	\\
HIP 78555	&	4	&	10.4	&	55712.2	&	4.368	&	54.4	&	56005.2	&	4.375	&	54.1	&	40.0	\\
HIP 78555	&	5	&	10.5	&	55712.2	&	4.025	&	106.5	&	56005.2	&	4.023	&	106.1	&	37.4	\\
HIP 78963	&	1	&	9.7	&	55712.2	&	4.583	&	61.3	&	56005.2	&	4.573	&	61.4	&	13.5	\\
HIP 78977	&	1	&	8.7	&	55712.2	&	3.739	&	30.3	&	56006.2	&	3.744	&	30.1	&	2.5	\\
HIP 79258	&	1	&	9.5	&	55712.2	&	4.247	&	98.3	&	56005.2	&	4.250	&	98.2	&	30.6	\\
HIP 79369	&	1	&	10.1	&	55696.2	&	2.685	&	196.2	&	56005.5	&	2.655	&	195.6	&	5.8	\\
HIP 79516	&	1	&	8.6	&	55696.2	&	4.585	&	181.9	&	56005.2	&	4.560	&	181.7	&	51.5	\\
HIP 79516	&	2	&	9.0	&	55696.2	&	4.227	&	337.2	&	56005.2	&	4.247	&	337.7	&	56.3	\\
HIP 79516	&	3	&	8.1	&	55696.2	&	4.692	&	342.5	&	56005.2	&	4.719	&	342.9	&	40.7	\\
HIP 79673	&	1	&	8.1	&	55696.2	&	2.692	&	255.4	&	56005.5	&	2.674	&	255.9	&	8.5	\\
HIP 79673	&	2	&	9.7	&	55696.2	&	2.561	&	132.5	&	56005.5	&	2.548	&	131.9	&	23.8	\\
HIP 79673	&	3	&	10.1	&	55696.2	&	2.346	&	147.6	&	56005.5	&	2.335	&	146.9	&	26.6	\\
HIP 79673	&	4	&	10.4	&	55696.2	&	2.696	&	285.7	&	56005.5	&	2.694	&	286.4	&	40.1	\\
HIP 79673	&	5	&	10.5	&	55696.2	&	3.374	&	103.8	&	56005.5	&	3.393	&	103.4	&	58.0	\\
HIP 79710	&	1	&	9.2	&	55696.2	&	2.797	&	282.0	&	56020.3	&	2.797	&	282.9	&	67.8	\\
HIP 79710	&	2	&	9.7	&	55696.2	&	2.877	&	126.8	&	56020.3	&	2.880	&	126.2	&	82.9	\\
HIP 79710	&	3	&	9.5	&	55696.4	&	3.463	&	128.2	&	56020.3	&	3.460	&	127.7	&	88.8	\\
HIP 79710	&	4	&	8.1	&	55696.4	&	4.106	&	54.2	&	56020.3	&	4.130	&	54.0	&	64.8	\\
HIP 79710	&	5	&	10.3	&	55696.4	&	3.720	&	161.1	&	56020.3	&	3.691	&	160.7	&	99.1	\\
HIP 79710	&	6	&	10.1	&	55696.4	&	4.862	&	73.4	&	56020.3	&	4.871	&	73.3	&	99.9	\\
HIP 79710	&	7	&	10.3	&	55696.4	&	4.836	&	82.7	&	56020.3	&	4.841	&	82.5	&	100.0	\\
HIP 79710	&	8	&	5.3	&	55696.4	&	3.898	&	73.6	&	56020.3	&	3.920	&	73.3	&	10.3	\\
HIP 79742	&	1	&	9.8	&	55675.5	&	2.658	&	293.2	&	56005.5	&	2.656	&	293.7	&	23.3	\\
HIP 80130	&	1	&	7.1$^{\rm c}$	&	55675.5	&	4.121	&	111.0	&	56338.5	&	4.117	&	111.0	&	1.0	\\
HIP 81851	&	1	&	8.9	&	55675.5	&	1.632	&	268.3	&	56020.3	&	1.614	&	269.2	&	2.4	\\
HIP 81851	&	2	&	10.3	&	55675.5	&	2.525	&	4.6	&	56020.3	&	2.553	&	4.9	&	15.5	\\
HIP 82534	&	1	&	10.4	&	55641.5	&	3.238	&	273.2	&	55989.5	&	3.240	&	273.5	&	36.1	\\
HIP 82534	&	2	&	10.4	&	55641.5	&	3.135	&	51.4	&	55989.5	&	3.151	&	51.2	&	34.3	\\
HIP 82534	&	3	&	9.5	&	55641.5	&	4.718	&	38.8	&	55989.5	&	4.726	&	38.7	&	37.6	\\
HIP 82569	&	1	&	8.0	&	55695.2	&	4.991	&	331.2	&	56336.5	&	5.010	&	331.2	&	52.8	\\
HIP 82569	&	2	&	8.5	&	55695.2	&	3.109	&	334.7	&	56336.5	&	3.119	&	334.8	&	35.0	\\
HIP 82569	&	3	&	9.6	&	55695.2	&	2.901	&	8.0	&	56336.5	&	2.908	&	7.8	&	58.8	\\
HIP 82569	&	4	&	10.2	&	55695.2	&	3.397	&	40.6	&	56336.5	&	3.393	&	40.5	&	85.7	\\
HIP 82569	&	5	&	10.6	&	55695.2	&	4.419	&	75.9	&	56336.5	&	4.414	&	76.0	&	98.9	\\
HIP 82569	&	6	&	9.8	&	55695.2	&	2.571	&	278.5	&	56336.5	&	2.572	&	278.7	&	55.7	\\
HIP 82569	&	7	&	9.5	&	55695.2	&	1.594	&	201.2	&	56336.5	&	1.603	&	201.0	&	21.9	\\
HIP 83159	&	1	&	9.9	&	55695.2	&	3.016	&	87.1	&	56012.2	&	3.023	&	86.9	&	86.6	\\
HIP 83159	&	2	&	9.2	&	55695.2	&	2.891	&	276.6	&	56012.2	&	2.892	&	277.1	&	65.6	\\
HIP 83159	&	3	&	10.3	&	55695.2	&	4.541	&	323.5	&	56012.2	&	4.560	&	323.7	&	99.8	\\
HIP 83159	&	4	&	10.3	&	55695.2	&	4.494	&	277.4	&	56012.2	&	4.491	&	277.7	&	99.8	\\
\label{t:astro}
\end{longtable}
\twocolumn

\begin{table*}[p]
\caption{Properties of the discovered systems.}
\label{t:binary}
\begin{tabular}{lccccccccc}
\hline
\hline
Target & Pair & Sep. & PA & $\Delta K$ & $a_{\rm proj}$ & $m_{\rm A}$ & $m_{\rm B}$ & $q$ & FAP \\
 & & (\arcsec) & (deg) & (mag) & (AU) & ($M_{\rm sun}$) & ($M_{\rm sun}$) & & \\
\hline
HIP 57238	&	AB	&	1.180$\pm$0.008	&	264.6$\pm$0.4	&	3.8$\pm$0.2	&	204	&	1.47	&	0.14	&	0.09	&	$4 \times 10^{-4}$	\\
HIP 57595	&	AB	&	0.142$\pm$0.004	&	251.3$\pm$0.9	&	1.3$\pm$0.2	&	24	&	1.49	&	0.99	&	0.66	&	$8 \times 10^{-7}$	\\
HIP 57710	&	AB	&	1.170$\pm$0.008	&	264.9$\pm$0.4	&	3.7$\pm$0.2	&	139	&	1.42	&	0.12	&	0.08	&	$1 \times 10^{-4}$	\\
HIP 58220	&	AB	&	0.760$\pm$0.008	&	315.9$\pm$0.4	&	3.6$\pm$0.2	&	80	&	1.44	&	0.14	&	0.10	&	$2 \times 10^{-4}$	\\
HIP 58528	&	AB	&	4.460$\pm$0.008	&	161.3$\pm$0.4	&	2.3$\pm$0.2	&	399	&	1.33	&	0.30	&	0.23	&	$2 \times 10^{-3}$	\\
HIP 58899	&	AB	&	4.218$\pm$0.008	&	258.0$\pm$0.4	&	2.5$\pm$0.2	&	483	&	1.49	&	0.44	&	0.29	&	$9 \times 10^{-4}$	\\
HIP 58899	&	AC	&	0.263$\pm$0.004	&	233.6$\pm$0.9	&	3.9$\pm$0.2	&	30	&	1.49	&	0.12	&	0.08	&	$1 \times 10^{-5}$	\\
HIP 59693	&	AB	&	0.427$\pm$0.004	&	170.6$\pm$0.9	&	1.8$\pm$0.2	&	51	&	1.21	&	0.37	&	0.30	&	$6 \times 10^{-6}$	\\
HIP 60885	&	AB	&	0.894$\pm$0.008	&	317.3$\pm$0.4	&	3.7$\pm$0.2	&	127	&	2.47	&	0.24	&	0.10	&	$1 \times 10^{-4}$	\\
HIP 62032	&	AB	&	0.309$\pm$0.004	&	127.5$\pm$0.9	&	2.6$\pm$0.2	&	52	&	1.82	&	0.51	&	0.28	&	$6 \times 10^{-6}$	\\
HIP 63272	&	AB	&	0.290$\pm$0.004	&	167.9$\pm$0.9	&	2.5$\pm$0.2	&	32	&	1.45	&	0.38	&	0.26	&	$4 \times 10^{-6}$	\\
HIP 65423	&	AB	&	1.835$\pm$0.005	&	247.4$\pm$0.2	&	4.4$\pm$0.1	&	228	&	1.10	&	0.07	&	0.06	&	$2 \times 10^{-3}$	\\
HIP 65517	&	AB	&	0.350$\pm$0.005	&	321.7$\pm$1.4	&	4.5$\pm$0.1	&	39	&	1.20	&	0.06	&	0.05	&	$3 \times 10^{-5}$	\\
HIP 67428	&	AB	&	3.563$\pm$0.008	&	327.0$\pm$0.4	&	3.8$\pm$0.2	&	416	&	1.44	&	0.11	&	0.08	&	$3 \times 10^{-3}$	\\
HIP 69291	&	AB	&	1.477$\pm$0.008	&	334.0$\pm$0.4	&	3.8$\pm$0.2	&	207	&	1.52	&	0.15	&	0.10	&	$1 \times 10^{-4}$	\\
HIP 71708	&	AB	&	3.450$\pm$0.008	&	73.0$\pm$0.4	&	2.3$\pm$0.2	&	445	&	1.45	&	0.44	&	0.30	&	$4 \times 10^{-4}$	\\
HIP 71767	&	AB	&	0.382$\pm$0.004	&	313.4$\pm$0.9	&	2.0$\pm$0.2	&	74	&	2.71	&	1.04	&	0.38	&	$4 \times 10^{-6}$	\\
HIP 72099	&	AB	&	0.667$\pm$0.005	&	34.4$\pm$0.4	&	4.2$\pm$0.1	&	107	&	1.40	&	0.10	&	0.07	&	$6 \times 10^{-5}$	\\
HIP 74104	&	AB	&	1.847$\pm$0.008	&	210.7$\pm$0.4	&	2.9$\pm$0.2	&	310	&	2.45	&	0.49	&	0.20	&	$3 \times 10^{-4}$	\\
HIP 75891	&	AB	&	0.437$\pm$0.004	&	314.3$\pm$0.9	&	2.5$\pm$0.2	&	68	&	2.36	&	0.61	&	0.26	&	$1 \times 10^{-5}$	\\
HIP 77038	&	AB	&	1.439$\pm$0.008	&	232.4$\pm$0.4	&	3.3$\pm$0.2	&	185	&	1.42	&	0.16	&	0.12	&	$2 \times 10^{-4}$	\\
HIP 77038	&	AC	&	1.426$\pm$0.008	&	243.2$\pm$0.4	&	4.2$\pm$0.2	&	183	&	1.42	&	0.09	&	0.07	&	$4 \times 10^{-4}$	\\
HIP 77388	&	AB	&	1.223$\pm$0.008	&	13.5$\pm$0.4	&	3.0$\pm$0.2	&	167	&	1.84	&	0.39	&	0.21	&	$8 \times 10^{-5}$	\\
HIP 77520	&	AB	&	2.217$\pm$0.008	&	196.7$\pm$0.4	&	1.9$\pm$0.2	&	291	&	1.41	&	0.50	&	0.36	&	$3 \times 10^{-4}$	\\
HIP 77815	&	AB	&	0.463$\pm$0.004	&	218.4$\pm$0.9	&	2.3$\pm$0.2	&	66	&	1.75	&	0.90	&	0.51	&	$4 \times 10^{-6}$	\\
HIP 78150	&	AB	&	1.631$\pm$0.008	&	111.0$\pm$0.4	&	2.1$\pm$0.2	&	238	&	2.65	&	0.92	&	0.35	&	$7 \times 10^{-4}$	\\
HIP 79054	&	AB	&	0.313$\pm$0.004	&	106.4$\pm$0.9	&	2.3$\pm$0.2	&	39	&	1.31	&	0.71	&	0.54	&	$3 \times 10^{-6}$	\\
HIP 79097	&	AB	&	0.814$\pm$0.008	&	340.0$\pm$0.4	&	3.3$\pm$0.2	&	139	&	1.99	&	0.75	&	0.38	&	$3 \times 10^{-5}$	\\
HIP 79392	&	AB	&	3.650$\pm$0.008	&	128.8$\pm$0.4	&	4.3$\pm$0.2	&	709	&	1.98	&	0.56	&	0.28	&	$1 \times 10^{-3}$	\\
\hline
\end{tabular}
\end{table*}

\end{document}